\documentclass[a4paper,fleqn,usenatbib]{mnras}
\usepackage{newtxtext,newtxmath}
\usepackage[T1]{fontenc}
\usepackage{ae,aecompl}
\usepackage{graphicx}	
\usepackage{amsmath}	
\usepackage{amssymb}	

\title[Carbon stars with increased oxygen and nitrogen abundances]{Carbon stars with
increased oxygen and nitrogen abundances: hydrostatic dust-free model atmospheres}

\author[B. Aringer et al.]{B.~Aringer,$^{1,3}$\thanks{E-mail: bernhard.aringer@unipd.it}
P.~Marigo,$^{1}$
W.~Nowotny,$^{3}$
L.~Girardi,$^{2}$
M.~Me\v{c}ina,$^{3}$
and A.~Nanni,$^{1,4}$\\
$^{1}$Dipartimento di Fisica e Astronomia Galileo Galilei,
Universit\`a di Padova, Vicolo dell'Osservatorio 3, I-35122 Padova, Italy\\
$^{2}$Osservatorio Astronomico di Padova -- INAF,
Vicolo dell'Osservatorio 5, I-35122 Padova, Italy\\
$^{3}$Department of Astrophysics, University of Vienna,
T\"urkenschanzstra{\ss}e 17, A-1180 Wien, Austria\\
$^{4}$Aix Marseille Univ., CNRS, CNES, LAM, Marseille, France}

\date{Accepted 2019 May 8. Received 2019 April 28; in original form 2019 March 13}
\pubyear{2019}

\begin{document}
\label{firstpage}
\pagerange{\pageref{firstpage}--\pageref{lastpage}}
\maketitle

\begin{abstract}
We have computed a grid of hydrostatic spherical {\sc COMARCS} models for
C stars covering metallicities from [Z/H]~=~0 to $-2$ and values of the
carbon excess [C$-$O] from 6.41 to 9.15, plus some temperature sequences, where
the amount of oxygen and nitrogen is increased relative to a scaled solar element
mixture. Such abundance variations may appear during the late stages of stellar
evolution. Our study covers changes of [O/Z] and [N/Z] going up to +0.5. Based on
the atmospheric structures we have calculated synthetic spectra and photometry for
all of the models in a consistent way.

The sequences with changed [O/Z] and [N/Z] can be used to derive correction
terms, which are applied to the colours predicted for a certain combination
of effective temperature, surface gravity, metallicity and carbon excess. If one
neglects these shifts in case of a variable oxygen amount, taking [C$-$O] instead
of C/O gives much better results, since the first quantity dominates the formation
of many important molecular species. For the warmer C giants with weaker pulsation
it is in principle possible to determine [C$-$O], [O/Z] or [N/Z] from high
resolution spectra, when the opacities in the radiative transfer calculations
for the models and observable properties are treated consistently. The
corresponding changes due to the abundances often become significantly larger
than the deviations caused by uncertainties of the stellar parameters or by a
optically thin dust shell. Photometric data and low or medium resolution
spectra are not sufficient to derive the mentioned quantities.
\end{abstract}

\begin{keywords}
stars: late-type -- stars: AGB and post-AGB -- stars: atmospheres --
stars: evolution -- molecular data -- Hertzsprung-Russell and colour-magnitude diagrams
\end{keywords}

\section{Introduction}

After the exhaustion of helium in the core, low- and intermediate-mass stars with
initial masses in the approximate range $\rm 0.8~M_{\odot} \la M_i \la 7~M_{\odot}$
may experience several mixing episodes during the latest stage of their
evolution, which is called the thermally pulsing asymptotic giant branch phase
\citep[TP-AGB,][]{Herwig05}. These events, known as "third dredge-up" take place in
conjunction with the thermal pulses of the He-burning shell and enrich the convective
envelope with newly synthesized elements, in particular $^4$He, $^{12}$C, $^{22}$Ne, $^{25}$Mg
and heavy isotopes produced by the s-process \citep{KarakasLattanzio14}. It is universally
acknowledged that the corresponding surface enhancement explains the existence of intrinsic
carbon stars characterized by a photospheric C/O ratio larger than one.

In this work we study the effect of enhanced oxygen and nitrogen abundances on the
atmospheres and spectra of carbon stars. Concerning the second of the two elements
stellar evolution models predict that some enrichment of $^{14}$N takes place as a
result of the first dredge-up, which is a mixing episode happening, when an
object expands and moves towards its Hayashi line after the end of its time on the
main sequence \citep[e.g.][]{ForestiniCharbonnel97}. The amount of additional $^{14}$N
is expected to depend on mass and metallicity. For instance, following recent
{\sc PARSEC} \citep{2012MNRAS.427..127B} calculations for stars with a solar-like
starting composition (metal mass fraction $\rm X_Z$~=~0.014), after the first
dredge-up the surface nitrogen increases with the initial mass from
[N/Z]~$\simeq$~0.2 at $\rm M_i = 1.0~M_{\odot}$ to [N/Z]~$\simeq$~0.55 at
$\rm M_i = 3.4~M_{\odot}$. A further enrichment may appear in intermediate-mass
objects with $\rm M_i > 3.5-4.0~M_{\odot}$ caused by the occurrence of the second
dredge-up at the base of the Early-AGB, when the H-burning shell is temporarily
extinguished. For $\rm X_Z$~=~0.014 the [N/Z] value in these giants can grow up to
$\simeq 0.65$ at $\rm M_i = 6.0~M_{\odot}$. In stars with lower metallicity the {\sc PARSEC}
models predict that the increase of nitrogen relative to its initial abundance
becomes a bit larger. For example, at $\rm X_Z$~=~0.001 [N/Z] is typically by
$\approx 0.05$ higher.

In current standard stellar evolution calculations usually no increase of the oxygen
abundance is predicted as a consequence of the first, second and third dredge-up
episodes. However, a possible enrichment of $^{16}$O may be expected during the
TP-AGB phase under particular conditions. Allowing for overshoot beyond the
Schwarzschild border of the pulse-driven convective zone during the third dredge-up, some
amount of primary $^{16}$O produced by the reaction $^{12}$C$(\alpha,\gamma)^{16}$O can be
mixed up to the surface \citep{Herwig00}. This scenario seems to be supported by abundance
analyses of PG1159 stars \citep[e.g.][]{WernerHerwig06}, post-AGB objects
\citep[e.g.][]{DeSmedtetal16} and planetary nebulae \citep[e.g.][]{Garciaetal16}.

Following earlier work by \citet{2000A&A...356..253J} and
\citet{2001A&A...371.1065L}, \citet{2009A&A...503..913A} have published a grid of
746 hydrostatic {\sc COMARCS} models for carbon star atmospheres, which covers effective
temperatures between 2400 and 4000~K with C/O ratios ranging from 1.05 to 5.0. The
included metallicities are [Z/H]~=~0, $-0.5$ and $-1$ (see the definition in
Section~\ref{sec_par}). The calculations, where spherical symmetry and local
thermodynamic plus chemical equilibrium (LTE) have been assumed, did not take
the formation of dust into account. Based on the models the authors computed
synthetic spectra and photometry, which could then be compared to observations. One
of the main conclusions from this work is that down to about 2800~K the {\sc COMARCS}
atmospheres are able to predict the measured colours of carbon stars, while for
cooler objects the simulated results appear much too blue. Those large deviations
occur because of circumstellar dust shells and the effects of pulsation plus
massloss on the radial temperature-pressure structures.

Compared to the work of \citet{2009A&A...503..913A} the new {\sc COMARCS} models used
here have been calculated with updated molecular opacities applied already in
\citet{2016MNRAS.457.3611A}, where K and M stars are studied. Some information
concerning these changes can be found in the next section. The impact on the
photometric results is discussed in Section~\ref{sec_discussion}. Our new {\sc COMARCS}
grid covers also a much larger set of abundance combinations including a very low
metallicity of [Z/H]~=~$-2$, C/O ratios from 1.01 to 10 and more, as well as
mixtures with an enrichment of oxygen and nitrogen. \citet{2017A&A...601A.141G}
compared medium to high resolution spectra of carbon stars taken with the X-shooter
instrument to synthetic ones based on the current models. They arrive at a
conclusion similar to the results of \citet{2009A&A...503..913A}. While the
hydrostatic calculations describe the warmer objects with weak pulsations quite
well, the cooler sources are dominated by dust and dynamic effects on the
structures.\footnote{Many of the carbon star models used here were already included
in the original database connected to \citet{2016MNRAS.457.3611A}. However, we did
not discuss them in that paper.}

Following the investigations of \citet{2011A&A...529A.129N,2013A&A...552A..20N}, a grid
of dynamic models for cool carbon giants with solar metallicity has been produced by
\citet{2014A&A...566A..95E}, who also published a database containing spectra and
photometric properties as a function of time. These calculations consider
pulsations plus the formation and opacity of dust, which may drive intense stellar
winds. Thus, they can explain the very red colours connected to high massloss
rates. The models were computed with a version of the {\sc DARWIN} code described in
\citet{2016A&A...594A.108H}. A setup similar to the one for the {\sc COMARCS} atmospheres
has been used to determine the observable properties. The authors show that their
approach reproduces the measured colours and outflow velocities of cool pulsating
carbon stars quite well. The main problem of those models is the limited capability
to describe objects with weak massloss and reddening, which are very common. {\sc DARWIN}
calculations also covering lower metallicities have recently been published by
\citet{2019A&A...623A.119B}.

Since the circumstellar reddening causes in pulsating carbon stars having massloss
by far the strongest effect on the overall spectral energy distribution, it is
a very common approach to determine the colours by combining a central hydrostatic
atmosphere with an extra description of the dust shell. Compared to the dynamic
calculations this procedure takes much less time, which is an advantage, if big
grids are needed. In addition, also the large number of objects generating a weak
wind can be reproduced without any problems. On the other hand, such models do not
consider temporal variations resulting in complicated structures, and using them
introduces free parameters. For example, the massloss rate or something equivalent
has to be assumed in all cases. One approach is the computation of stationary
winds including dust formation and opacity like in \citet{2015MNRAS.447.2992D} or
\citet{2016MNRAS.462.1215N}. Shell models based completely on parameters
except for the radiative transfer were used, for instance, by
\citet{2011A&A...532A..54S}, \citet{2012A&A...543A..36G}, \citet{2012A&A...537A.105G}
and \citet{2012ApJ...753...71R}. The method of applying circumstellar reddening to
a central hydrostatic atmosphere is also the standard tool for the determination of
colours of cool giants in connection with population synthesis codes
\citep{2008A&A...482..883M,2017ApJ...835...77M,2019MNRAS.485.5666P}.

\section{Model Atmospheres and Synthetic Spectra}
\label{sec_mods}

In order to simulate the atmospheric structures of carbon stars we use the newest
hydrostatic models from the {\sc COMARCS} grid described in \citet{2016MNRAS.457.3611A}
where one can find more details concerning the calculations and the included
opacity sources. The corresponding database, which contains mainly objects of
spectral type K, M, S and C, was significantly extended in the domain with
C/O~$>$~1 for the purpose of the investigation presented here. An updated list
of the available stellar parameter and elemental abundance combinations is given
in \url{http://stev.oapd.inaf.it/atm}. Concerning the atomic and molecular
transitions taken into account and the treatment of convection all models used
in this work have the same setup as in \citet{2016MNRAS.457.3611A}. The {\sc COMARCS}
program, where spherical symmetry and local thermodynamic and chemical equilibrium
(LTE) are assumed, was originally developed based on a version of the {\sc MARCS} code
\citep{1975A&A....42..407G,2008A&A...486..951G} described in
\citet{1992A&A...261..263J} and \citet{1997A&A...323..202A}. The formation
of dust is not included.

Compared to the previous {\sc COMARCS} grid of carbon stars published by
\citet{2009A&A...503..913A} some molecular opacities have been added and
updated as described in \citet{2016MNRAS.457.3611A}. For example, the CrH
bands are much weaker in the new calculations. As a consequence, their
effect on medium or low resolution spectra and photometry became almost
negligible. We also want to note that the scaling of the C$_2$
list from \citet{1974A&A....31..265Q} at wavelengths above
1.15~$\mu$m, which has been suggested by \citet{2001A&A...371.1065L}
and used as a standard for the 2009 models, is not applied anymore. It was
already found by \citet{2009A&A...503..913A} that it causes synthetic
(H$-$K) colours shifted by 0.1~mag with respect to the observations. At
least for the bluer objects with weak pulsations and low massloss rates
these systematic deviations disappear, if the original data are not
modified. Some of the photometric changes between the old and the current
carbon star grid will be discussed later in this work.

When calculating observable properties of our models, we have again followed
the approach described in \citet{2016MNRAS.457.3611A}. Synthetic medium
resolution R~=~10000 opacity sampling (OS) and convolved low resolution
R~=~200 spectra were produced with the {\sc COMA} code \citep{2000DissAri}, which
was already used to create gas absorption data for the generation of the
{\sc COMARCS} atmospheres. Based on the output of these computations we could
determine photometric results for a large number of filters covering
many different systems. Tables with the corresponding bolometric corrections
(BC) are available at \url{http://stev.oapd.inaf.it/atm}. This
database, which comprises the complete {\sc COMARCS} grid including all models
used here, also contains the low and medium resolution spectra. It will be
extended if further results are produced. We want to note that due to their
statistical nature the OS spectra should never be directly compared to
observations.

Concerning the photometric data we want to note that in the current work only
visual and near infrared magnitudes, which were determined according to the
definition from \citet{1990PASP..102.1181B} and \citet{1988PASP..100.1134B}, will
be discussed. The V, J, H and K values and corresponding colours appearing in
\ref{sec_phot} and \ref{sec_discussion} are based on this system (Bessell).

In addition to the standard approach, we have used {\sc COMA} to calculate high
resolution spectra with R~=~300000 covering the wavelength range between
1.7046 and 1.7064~$\mu$m. This was done only for a few selected models
having effective temperatures of 2800 and 3300~K\@. Some of the results
are shown in Fig.~\ref{cst_hires}.

\begin{table}
\centering
\caption{The opacity sets used for this work. The overall metallicity [Z/H] and
additional changes of the amount of oxygen [O/Z] or nitrogen [N/Z] are listed with
respect to the solar values. The elemental abundances of O, C and Fe correspond
to number densities: $\rm log(\varepsilon_{Fe,O,C}/\varepsilon_H) + 12$. Combinations
taken from the standard {\sc COMARCS} grid are marked with an "s".}
\label{ari_comarcs}
\begin{tabular}{rrcrrccl}
\hline
[Z/H] & C/O & [C$-$O] & [O/Z] & [N/Z] & O & C & Fe\\
\hline
-2.0 &  1.40 & 6.41 &  0.00 &  0.0 & 6.80 & 6.95 & 5.56$\rm ^s$\\
-2.0 &  2.00 & 6.80 &  0.00 &  0.0 & 6.80 & 7.10 & 5.56$\rm ^s$\\
-2.0 &  6.00 & 7.50 &  0.00 &  0.0 & 6.80 & 7.58 & 5.56$\rm ^s$\\
-2.0 & 10.00 & 7.76 &  0.00 &  0.0 & 6.80 & 7.80 & 5.56$\rm ^s$\\
-2.0 & 29.47 & 8.26 &  0.00 &  0.0 & 6.80 & 8.27 & 5.56\\
-2.0 & 10.00 & 8.26 & +0.50 &  0.0 & 7.30 & 8.30 & 5.56\\
-2.0 & 10.00 & 8.26 & +0.50 & +0.5 & 7.30 & 8.30 & 5.56\\
\hline
-1.0 &  1.40 & 7.36 & -0.04 &  0.0 & 7.76 & 7.91 & 6.56\\
-1.0 &  1.58 & 7.53 & -0.04 &  0.0 & 7.76 & 7.96 & 6.56\\
-1.0 &  2.00 & 7.76 & -0.04 &  0.0 & 7.76 & 8.06 & 6.56\\
-1.0 &  2.04 & 7.78 & -0.04 &  0.0 & 7.76 & 8.07 & 6.56\\
-1.0 &  2.95 & 8.05 & -0.04 &  0.0 & 7.76 & 8.23 & 6.56\\
-1.0 &  3.90 & 8.22 & -0.04 &  0.0 & 7.76 & 8.35 & 6.56\\
-1.0 &  1.10 & 6.80 &  0.00 &  0.0 & 7.80 & 7.84 & 6.56$\rm ^s$\\
-1.0 &  1.40 & 7.41 &  0.00 &  0.0 & 7.80 & 7.95 & 6.56$\rm ^s$\\
-1.0 &  2.00 & 7.80 &  0.00 &  0.0 & 7.80 & 8.10 & 6.56$\rm ^s$\\
-1.0 &  4.00 & 8.28 &  0.00 &  0.0 & 7.80 & 8.40 & 6.56$\rm ^s$\\
-1.0 &  8.00 & 8.65 &  0.00 &  0.0 & 7.80 & 8.71 & 6.56$\rm ^s$\\
-1.0 &  1.40 & 7.53 & +0.12 &  0.0 & 7.92 & 8.07 & 6.56\\
-1.0 &  2.00 & 8.05 & +0.25 &  0.0 & 8.05 & 8.35 & 6.56\\
\hline
-0.5 &  1.05 & 7.00 &  0.00 &  0.0 & 8.30 & 8.32 & 7.06$\rm ^s$\\
-0.5 &  1.10 & 7.30 &  0.00 &  0.0 & 8.30 & 8.34 & 7.06$\rm ^s$\\
-0.5 &  1.40 & 7.91 &  0.00 &  0.0 & 8.30 & 8.45 & 7.06$\rm ^s$\\
-0.5 &  2.00 & 8.30 &  0.00 &  0.0 & 8.30 & 8.60 & 7.06$\rm ^s$\\
-0.5 &  4.00 & 8.78 &  0.00 &  0.0 & 8.30 & 8.90 & 7.06$\rm ^s$\\
-0.5 &  8.00 & 9.15 &  0.00 &  0.0 & 8.30 & 9.21 & 7.06$\rm ^s$\\
\hline
 0.0 &  1.01 & 6.81 &  0.00 &  0.0 & 8.80 & 8.81 & 7.56$\rm ^s$\\
 0.0 &  1.05 & 7.50 &  0.00 &  0.0 & 8.80 & 8.82 & 7.56$\rm ^s$\\
 0.0 &  1.10 & 7.80 &  0.00 &  0.0 & 8.80 & 8.84 & 7.56$\rm ^s$\\
 0.0 &  1.32 & 8.30 &  0.00 &  0.0 & 8.80 & 8.92 & 7.56\\
 0.0 &  1.40 & 8.41 &  0.00 &  0.0 & 8.80 & 8.95 & 7.56$\rm ^s$\\
 0.0 &  2.00 & 8.80 &  0.00 &  0.0 & 8.80 & 9.10 & 7.56$\rm ^s$\\
 0.0 &  1.10 & 8.30 & +0.50 &  0.0 & 9.30 & 9.34 & 7.56\\
 0.0 &  1.10 & 8.30 & +0.50 & +0.5 & 9.30 & 9.34 & 7.56\\
\hline
\end{tabular}
\end{table}

\subsection{Model Parameters}
\label{sec_par}

Temperature sequences of {\sc COMARCS} carbon star models calculated with oxygen
and nitrogen abundances deviating from the standard setup of a scaled solar
composition, which is described in Section~\ref{sec_sta}, are only available for
$\rm log(g~[cm/s^2]) = 0.0$ and the mass of the Sun. Thus, the investigation
presented here remains restricted to this selection of the corresponding two
parameters. The effect of varying their values has been discussed in detail by
\citet{2009A&A...503..913A}. As was already mentioned before, some of the
opacities were revised since then, resulting in moderate differences concerning
spectra and colours. However, the relative changes due to the stellar properties
remain almost the same. Because the production of complete carbon star subgrids
containing many combinations of values for the effective temperature, surface
gravity and mass takes much time, they exist in the {\sc COMARCS} database only for
standard cases with scaled solar abundances (at the moment all sets with
[Z/H]~=~0 and some with $-0.5$). Their upper limit of $\rm log(g~[cm/s^2])$ is
usually 2.0, while the lower one ranges between 0.0 for $\rm T_{eff} > 3400~K$
and $-1$ for the coolest objects. In addition, masses from 1.0 to 3.0~M$_{\odot}$
are covered. Detailed information concerning the available {\sc COMARCS} models can
be found at the internet link mentioned above.

The $\rm log(g~[cm/s^2]) = 0.0$ sequences discussed here and the {\sc COMARCS} carbon
star subgrids usually cover the effective temperature range from 2500 to 4400 or
4000~K with a maximum step size of 100~K\@. For a few of the standard cases with
scaled solar abundances the lower limit is 2600~K (C/O~=~1.01 and 1.10 at
[Z/H]~=~0). Like in \citet{2009A&A...503..913A} the microturbulent velocity
was set to $\xi = 2.5$~km/s. Concerning the chemical composition of the
Sun, which represents the reference for all other mixtures used in this
work, we followed the assumptions in \citet{2016MNRAS.457.3611A} based on
\citet{2009MmSAI..80..643C,2009A&A...498..877C}. The resulting C/O ratio is
close to 0.55.

In Table~\ref{ari_comarcs} we list the various abundance sets available for
carbon stars. The overall metallicity is described by the parameter [Z/H], which
corresponds to the logarithm of the number density ($\varepsilon$) for the bulk
of the individual elements heavier than He divided by the value for H
($\rm \varepsilon_H$) and scaled relative to the solar ratio. Since we use
here the common definition, where $\rm log(\varepsilon_H)$ is set constant
to 12, and no {\sc COMARCS} models with special deviations of the Fe content
exist, the quantities [Z/H], [Fe/H] and log~(Z/Z$_{\odot}$) are always
equal. A variation of the nitrogen and oxygen abundance with respect to the
other metals does not change them. The corresponding number densities are
described in a similar way by the solar scaled logarithmic ratios [O/Z] and
[N/Z]. The last parameter needed to characterize our sets is C/O
($\rm \varepsilon_C/\varepsilon_O$), which has to be larger than one
in order to form a carbon star. Table~\ref{ari_comarcs} contains also
the individual abundances of O, C and Fe defined by
$\rm log(\varepsilon_{Fe,O,C}/\varepsilon_H) + 12$ (solar values: 8.80
for O, 8.54 for C and 7.56 for Fe). In addition, we list
the important quantity [C$-$O], which stands for the difference of the
number densities of those two elements. It gives an estimate for the
amount of free carbon atoms not bound in CO and has to be calculated by
$\rm log((\varepsilon_C - \varepsilon_O)/\varepsilon_H) + 12$.

\subsubsection{Standard Models with Scaled Solar Abundances}
\label{sec_sta}

In order to obtain the chemical standard mixtures for the {\sc COMARCS} grid
(marked with "s" in Table~\ref{ari_comarcs}) we have first scaled the abundances
of the elements heavier than He to the chosen overall metallicity [Z/H] as
explained above. Subsequently, the amount of carbon was changed according
to the selected C/O ratio. At the moment models with values of
1.01, 1.05, 1.10, 1.40, 2.0, 4.0, 6.0, 8.0 and 10.0 are available. The coverage
depends on [Z/H], since we have to take into account that for metal-poor stars
the increase of C/O following a dredge-up event is much larger. In these objects
one needs also a higher ratio to get a certain amount of free carbon atoms [C$-$O].

Our database includes also two $\rm log(g~[cm/s^2]) = 0.0$ sequences with scaled
solar abundances, which have the same [C$-$O] values as the models discussed
in the next section, where the amount of oxygen and nitrogen is increased. The
corresponding parameters are C/O~=~1.32, [C$-$O]~=~8.30 at [Z/H]~=~0.0 and
C/O~=~29.47, [C$-$O]~=~8.26 at [Z/H]~=~$-2.0$.

\subsubsection{Models with a Fixed Enhancement of Oxygen and Nitrogen}

In order to study the effect of deviations from a scaled solar chemical
mixture in carbon stars we have produced two classes of models, where
only oxygen or oxygen and nitrogen are enhanced by a fixed factor. The
corresponding values are [O/Z]~=~+0.5 and [O/Z]~=~[N/Z]~=~+0.5. This
investigation was done for two abundance sets from the standard {\sc COMARCS}
grid keeping the overall metallicity and the C/O ratio constant. They
have the following parameters: C/O~=~1.10 at [Z/H]~=~0.0 and C/O~=~10.0
at [Z/H]~=~$-2.0$. It should be noted that an enhancement of oxygen will
result in a larger amount of free carbon atoms, if C/O and [Z/H] remain
unchanged. However, as was already mentioned in the previous section, in
both cases discussed here we have also computed sequences with scaled solar
abundances and the same [C$-$O].

\subsubsection{Enhancement of Oxygen Based on Stellar Evolution Models}
\label{sec_evo}

We replaced the carbon and oxygen abundances in our standard {\sc COMARCS} sets
having [Z/H]~=~$-1$ with the ones taken from a TP-AGB stellar evolution
track, which was computed with the {\sc COLIBRI} code \citep{2013MNRAS.434..488M}
adopting [Fe/H]~=~$-1$ and an initial mass of 2.0~M$_{\odot}$. In order to take
the effect of an O enrichment into account, the following mass fractions of the
involved isotopes have been assumed for the intershell: X($^{12}$C)~=~0.48 and
X($^{16}$O)~=~0.17. These values are very close to the ones that match the
observed photospheric composition of some PG1159 stars
\citep[see][for a detailed discussion]{Herwig00}. The two stages, where the
C/O ratio reaches 1.4 and 2.0 as a result of the third dredge-up
episodes, have been selected to create sequences of synthetic atmospheres
and spectra. This corresponds to [O/Z] values of +0.12 and +0.25, which are
smaller than the fixed shifts discussed in the previous section.

We have also calculated {\sc COMARCS} atmospheres using the initial oxygen
abundance of the stellar evolution models taken from a phase before the
start of the third dredge-up. It has a value, which is quite close to the
one in our standard grid. We do not expect that the corresponding shift
of [O/Z]~=~$-0.04$ will cause any significant changes of the radial
temperature and pressure structures or the spectra. For both of the
sequences with an oxygen enhancement three classes of such {\sc COMARCS} models
were produced having the same C/O (1.4, 2.0), [C$-$O] (7.53, 8.05) and
absolute carbon abundance (8.07, 8.35).

\begin{figure}
\includegraphics[width=6.6cm,clip,angle=270]{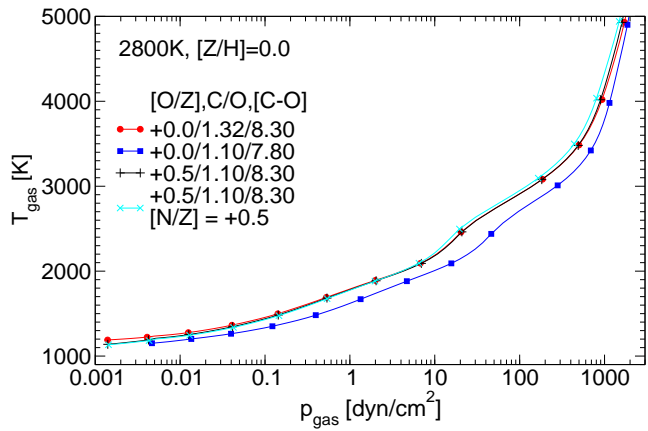}
\caption{The effect of an increased oxygen and nitrogen abundance on
the atmospheric temperature versus gas pressure structure is shown for
different {\sc COMARCS} models with $\rm T_{eff} = 2800~K$, $\rm log(g~[cm/s^2]) = 0.0$,
solar mass and metallicity. Results obtained adopting [O/Z]~=~+0.5 (black) and
[O/Z]~=~+0.5, [N/Z]~=~+0.5 (cyan) are compared to calculations for the standard
composition ([O/Z]~=~0, [N/Z]~=~0) with a constant C/O (blue) or [C$-$O] (red).
The plot symbols placed along the curves mark steps of
$\rm \Delta log(\tau_{Ross}) = 0.5$ starting with $-5$ at the outer edge.}
\label{cst_struc}
\end{figure}

\section{Results}
\label{sec_results}

In this section we discuss some of the important results from our
investigation concerning atmospheric structures, spectra with low or
high resolution and photometric colours.

\subsection{Model Structures}
\label{sec_struc}

In Fig.~\ref{cst_struc} we show the temperature pressure structures of four
{\sc COMARCS} models with $\rm T_{eff} = 2800~K$ and [Z/H]~=~0.0. The results, where
only oxygen or oxygen and nitrogen have been enhanced by [O/Z]~=~+0.5 and
[N/Z]~=~+0.5, are compared to calculations with scaled solar abundances and
the same C/O (1.10) or [C$-$O] (8.30). One can see that the atmospheres behave
in a very similar way, if the excess of free carbon atoms remains constant, while
the decrease of this quantity causes significant changes. For the lower value of
[C$-$O] the curve moves to cooler temperatures. Since the corresponding typical
shift ranges from 100 to 300~K, it has a considerable effect on the emerging
spectra. The main reason for this variation of the structure is that the amount
of free carbon influences the formation of species like C$_2$, C$_3$, HCN
and C$_2$H$_2$, which are important for the overall opacity.

If the stars get much warmer than about 3000 to 3200~K, the role of [C$-$O] as
a dominant parameter becomes less pronounced. This is due to the fact that the
abundance and importance of the mentioned molecules decreases with a higher
temperature. In the hotter models also variations of [O/Z] at a constant carbon
excess and of [N/Z] can have a moderate effect on the structures.

\begin{figure}
\includegraphics[width=6.5cm,clip,angle=270]{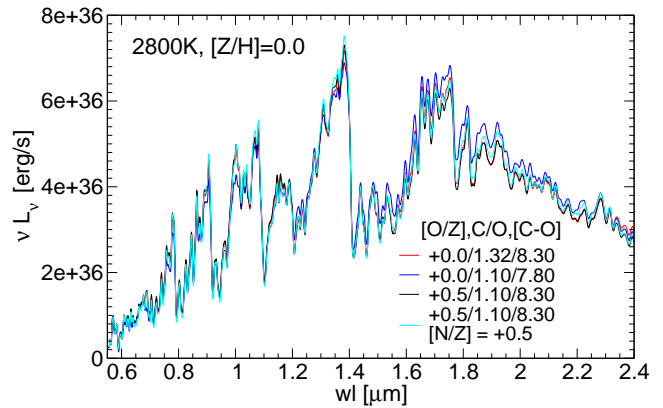}
\caption{Low resolution R~=~200 spectra computed from {\sc COMARCS} models
with $\rm T_{eff} = 2800~K$, $\rm log(g~[cm/s^2]) = 0.0$,
solar mass and metallicity. The effect of an increased oxygen and nitrogen
abundance is shown. Results obtained adopting [O/Z]~=~+0.5 (black) and
[O/Z]~=~+0.5, [N/Z]~=~+0.5 (cyan) are compared to calculations for the standard
composition ([O/Z]~=~0, [N/Z]~=~0) with a constant C/O (blue) or [C$-$O] (red).}
\label{cst_spec01}
\end{figure}

\begin{figure}
\includegraphics[width=7.1cm,clip,angle=270]{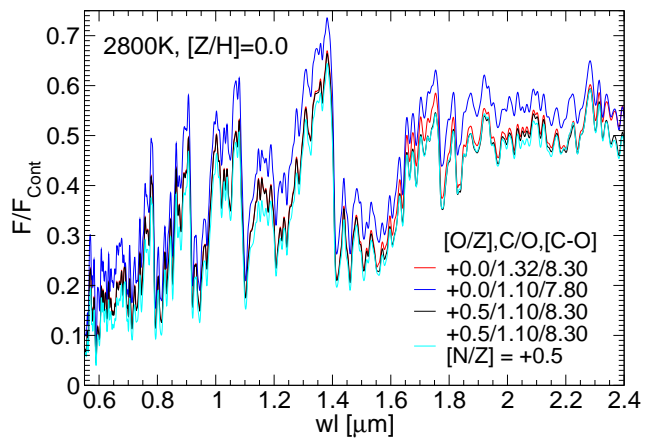}
\caption{Continuum normalized low resolution R~=~200 spectra based on {\sc COMARCS}
atmospheres. The included models and the corresponding line colours are the same
as in Fig.~\ref{cst_spec01}.}
\label{cst_spec02}
\end{figure}

\subsection{Low Resolution Spectra}
\label{sec_specl}

In Figs.~\ref{cst_spec01} and \ref{cst_spec02} we present low resolution spectra
covering the range from 0.6 to 2.4~$\mu$m, which have been computed for the four
{\sc COMARCS} models with $\rm T_{eff} = 2800~K$ and [Z/H]~=~0.0 used already in
Fig.~\ref{cst_struc}. This allows us to study the effects of an enrichment
of oxygen and oxygen plus nitrogen, if the C/O or [C$-$O] value remains
constant. The results shown in Fig.~\ref{cst_spec02} are normalized with respect
to a radiative transfer calculation, where all line opacities have been set to
zero. Using this inconsistent approach one can estimate the intensity of the
atomic and molecular absorption produced in different regions of the stellar
spectra without the influence of changes caused by the continuum.

Looking at Fig.~\ref{cst_spec02} we can conclude that the {\sc COMARCS} atmosphere
with the lower [C$-$O] value (7.80) has in general clearly the weakest line
absorption, while the strongest one is usually created by the model having
an increased amount of oxygen and nitrogen. For the cool effective temperature
of the selected examples and the shown spectral range the opacity is dominated
by molecular transitions of CN, C$_2$, C$_3$, HCN and C$_2$H$_2$. Atomic features
do not play an important role. Thus, the mentioned behaviour agrees with the
expectations, since a smaller amount of free carbon will decrease the formation
of the species listed above. In addition, CN and HCN may become more abundant, if
a larger fraction of nitrogen is available. However, the relation could be less
simple, because the opacity of the discussed molecules changes the atmospheric
structure, which influences then again the chemical equilibrium. One should also
take into account that in cool objects it is not possible to measure the
absorption relative to an absolute continuum, since even at higher resolution
all regions of the spectra are significantly affected by many overlapping
lines. The observable depth of certain features with respect to their
surrounding can often show a behaviour, which does not reflect the variation
of the total intensity of the corresponding molecular bands.

The unscaled spectra and deduced photometric data will be affected by millions
of weak overlapping lines as well as by the level and shape of the continuum
depending on the atmospheric structure. Despite the different intensity of their
molecular absorption, especially when the [C$-$O] value is varied, all results
shown in Fig.~\ref{cst_spec01} look quite similar. The general agreement of the
overall fluxes can be explained with the constant integrated luminosity of the
consistent calculations. But also on the smaller scales we find in many regions
only weak or no changes. This is at least partly due to a saturation of the
strong lines and the large influence of the molecular opacities on the
structures. Some moderate variations appear in the range around
1.8~$\mu$m, where a lower excess of free carbon atoms decreases the intensity of
the features. However, in contrast to the normalized data a broader spectral
interval with a clear positive correlation between the studied abundances and
the depth of the bands does not exist.

For a warmer star with an effective temperature of 3300~K the behaviour
is similar. In the normalized spectra the overall molecular absorption
clearly grows with a higher carbon excess or nitrogen abundance. The
variation of the unscaled results remains again much smaller. Nevertheless, the
correlation of the depth with the [C$-$O] and [N/Z] values is for many of the
features there more pronounced than at 2800~K.

\subsection{High Resolution Spectra}
\label{sec_spech}

\begin{figure*}
\includegraphics[width=8.8cm,clip,angle=0]{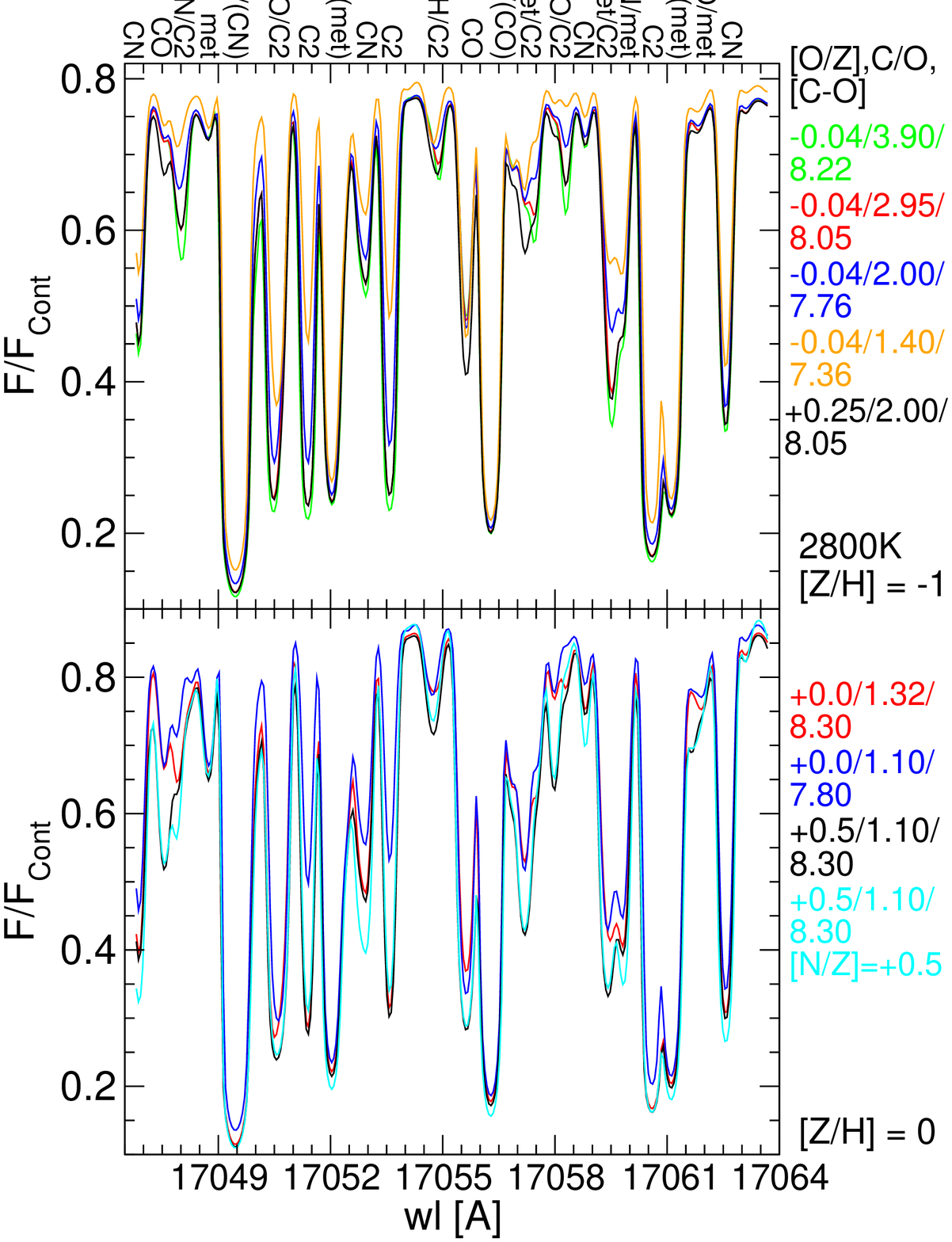}
\includegraphics[width=8.8cm,clip,angle=0]{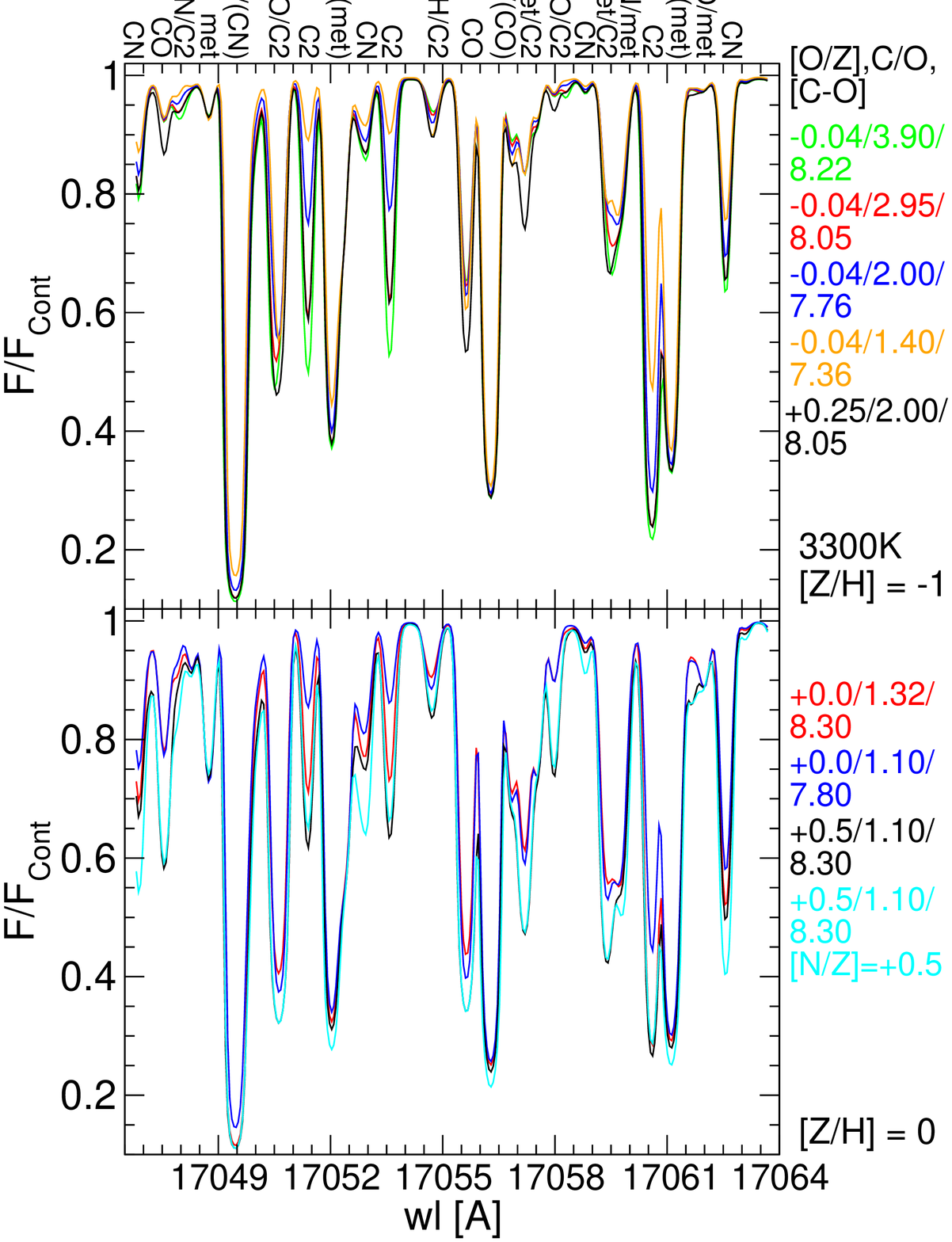}
\caption{High resolution R~=~300000 spectra based on {\sc COMARCS} models with
$\rm log(g~[cm/s^2]) = 0.0$ and one solar mass having $\rm T_{eff} = 2800~K$
(left panels) or 3300~K (right panels) and [Z/H]~=~0 (lower panels) or $-1$
(upper panels). The data are normalized relative to a calculation without
molecular and atomic line opacities. The different absorption features have
been marked with the species creating them. "met" stands for transitions due to
metals. If a certain component of a blend is much weaker than the rest, it was
put in brackets. For [Z/H]~=~$-1$ we compare an oxygen enriched model
([O/Z]~=~0.25, C/O~=~2.0, [C$-$O]~=~8.05, black) to calculations with
[O/Z]~=~$-0.04$ having a constant C/O ([C$-$O]~=~7.76, blue), [C$-$O]
(C/O~=~2.95, red), total amount of carbon $\rm \varepsilon_C$
(C/O~=~3.90, [C$-$O]~=~8.22, green) and a lower C/O of 1.4
([C$-$O]~=~7.36, orange). For [Z/H]~=~0 the shown abundance combinations
and the corresponding line colours are the same as in Fig.~\ref{cst_spec01}.}
\label{cst_hires}
\end{figure*}

The different panels in Fig.~\ref{cst_hires} contain high resolution spectra
based on {\sc COMARCS} models with an effective temperature of 2800 or 3300~K and
a metallicity of [Z/H]~=~0 or $-1$. The shown wavelength range between 1.7047
and 1.7064~$\mu$m is a typical example for a region characterized by a moderate
molecular absorption. Like in Fig.~\ref{cst_spec02} the data are normalized
relative to a calculation, where all line opacities have been omitted. Compared
to the unscaled results this makes apart from the general flux level no
difference, since the continuum does not change within such a narrow
interval. However, the shown spectra of the cooler models demonstrate clearly
that for the entire frequency range a large fraction of the radiation is
blocked by many weak overlapping molecular transitions. Even at a high
resolution there exists no point, which is not affected by strong line
absorption. This agrees with the results found in the previous section. The
situation changes, if the atmospheres become warmer. In Fig.~\ref{cst_hires}
one can see that the spectra computed from the models having 3300~K include
already some narrow regions, where the flux is close to the continuum level.

For the solar metallicity Fig.~\ref{cst_hires} contains the abundance combinations
already used in Figs.~\ref{cst_struc}, \ref{cst_spec01} and \ref{cst_spec02}. This
shows the spectral variation with changed values of [O/Z] and [O/Z] plus [N/Z], if
C/O or [C$-$O] is kept constant. In the case of [Z/H]~=~$-1$ we present
results, where the oxygen enrichment has been taken from the stellar evolution
calculations. {\sc COMARCS} models with [O/Z]~=~0.25 and C/O~=~2.0 are compared to
similar ones having [O/Z]~=~$-0.04$ and the same C/O, [C$-$O] (8.05) or total
amount of carbon (8.35). In addition, we include spectra for a much lower
C/O ratio of 1.4.

The intensities of the spectral features, which can be uniquely assigned to certain
molecules, are usually well correlated with the abundances of the involved
elements. The depth of the C$_2$ lines depends for example mainly on the carbon
excess. It increases, if the [C$-$O] values listed in the legends of the panels
become larger. However, for the strongest transitions the mentioned behaviour
is less pronounced, since they are saturated showing only weak variations due
to the chemical mixture. In the warmer models with solar metallicity some of the
C$_2$ lines also seem to be affected by an additional parameter possibly linked to
[O/Z]. The intensity of the CN features increases, when the values of [C$-$O], [N/Z]
and [Z/H] grow. It should be noted that the combination of the last two quantities
corresponds to the absolute nitrogen abundance. For the strong CN transitions we
find only weak or no changes caused by different chemical mixtures, which is again
connected to saturation. In some of these cases a higher [N/Z] results in slightly
deeper lines. Finally, the intensity of the CO features increases for larger values
of [O/Z] and [Z/H], where more oxygen is available.

A more complex relation between strength and elemental abundances can be found
for blends containing at least two transitions of different species. The behaviour
may change with effective temperature and metallicity, if these parameters influence
the intensity ratio of the components. A good example is the feature at
1.70506~$\mu$m, which consists of a C$_2$ and a CO line. In the cooler models
with [Z/H]~=~$-1$ the depth of this blend depends mainly on the carbon
excess. We see a clear increase for larger [C$-$O] values, which has also been
observed in the case of pure C$_2$ transitions. On the other hand, in the warmer
{\sc COMARCS} atmospheres having solar metallicity the behaviour resembles the one
of a CO line, where the oxygen abundance is the dominant parameter. For the
remaining two combinations of effective temperature and [Z/H] shown in
Fig.~\ref{cst_hires} [C$-$O] and [O/Z] both play an important role.

Another interesting blend can be observed around 1.70548~$\mu$m. It comprises two
lines produced by C$_2$ and by OH, which is a very rare species in the atmospheres
of carbon stars. However, the appearance of such OH features shows that even in
chemical equilibrium a small fraction of the oxygen is not bound in CO molecules. The
corresponding excess will increase, if the absolute abundance of the element grows
due to a higher [O/Z] or [Z/H], and for lower [C$-$O] values. In the models having
solar metallicity the intensity of the discussed blend changes according to these
predictions. This means that the OH transition is the dominant component. For
[Z/H]~=~$-1$ the behaviour becomes more complex, because the C$_2$ line, which
gets stronger with a larger [C$-$O], plays a role.

\subsection{Synthetic Photometry}
\label{sec_phot}

\begin{figure}
\includegraphics[width=9.9cm,clip,angle=0]{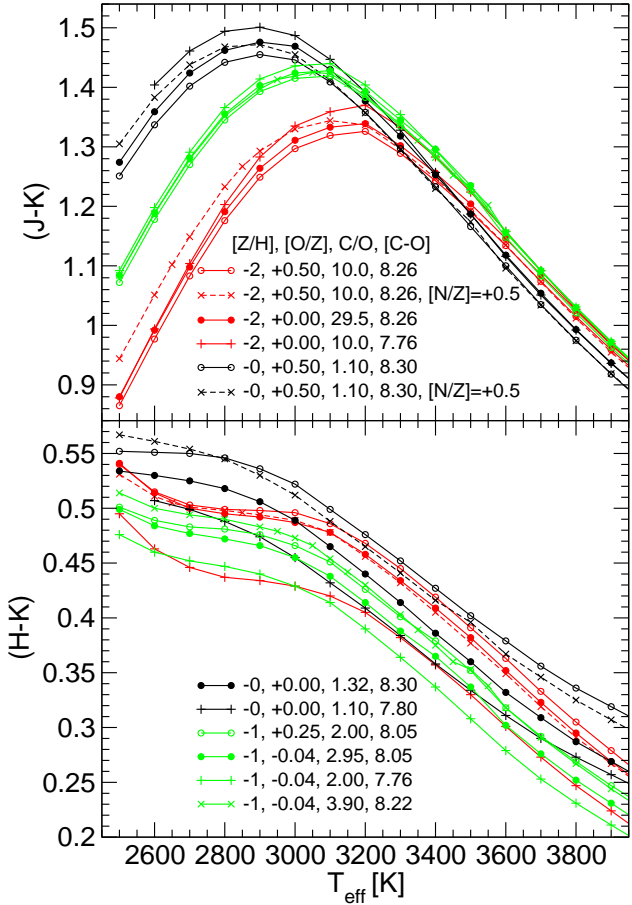}
\caption{(J$-$K) and (H$-$K) as a function of the effective temperature for
{\sc COMARCS} models with $\rm log(g~[cm/s^2]) = 0.0$ and one solar mass. The colours
correspond to different metallicities. For [Z/H]~=~0 (black) and $-2$ (red) results
obtained adopting [O/Z]~=~+0.5 (open circles) and [O/Z]~=~+0.5, [N/Z]~=~+0.5
(x, dashed lines) are compared to calculations for the standard composition
([O/Z]~=~0, [N/Z]~=~0) with a constant C/O (plus) or [C$-$O] (filled circles). In the
case of [Z/H]~=~$-1$ (green) we show oxygen enriched models having [O/Z]~=~0.25,
C/O~=~2.0 and [C$-$O]~=~8.05 (open circles) together with sequences where [O/Z] is
set to $-0.04$. They were created with the same C/O (plus), [C$-$O] (filled circles)
and total amount of carbon $\rm \varepsilon_C$ (x).}
\label{cst_colt}
\end{figure}

\begin{figure*}
\includegraphics[width=13.5cm,clip,angle=270]{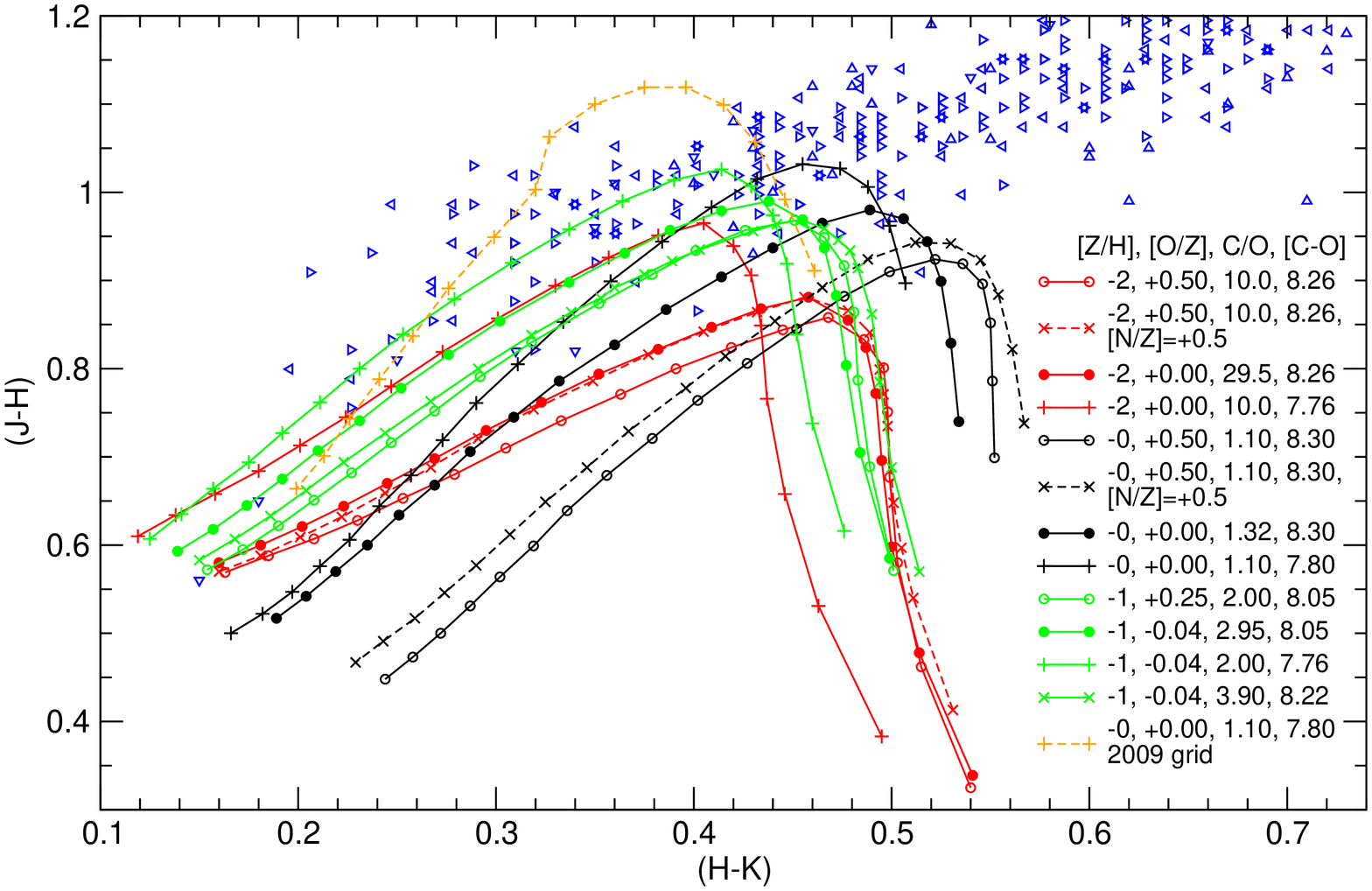}
\caption{(J$-$H) versus (H$-$K) for temperature sequences of {\sc COMARCS} models with
$\rm log(g~[cm/s^2]) = 0.0$ and one solar mass. The shown abundance combinations
and the corresponding colours and symbols are the same as in Fig.~\ref{cst_colt}.
In addition, we present results with [Z/H]~=~0.0 and C/O~=~1.10 taken from the
previous {\sc COMARCS} carbon star grid by \citet{2009A&A...503..913A}, where the mass
and surface gravity are identical (orange crosses, dashed line). The models are
compared to observational data (blue triangles) listed in
\citet[][up]{2001A&A...369..178B}, \citet[][left]{1981ApJ...249..481C},
\citet[][right]{1996AJ....112.2607C} and \citet[][down]{2016A&A...589A..36G}.}
\label{cst_colcol}
\end{figure*}

In the following text we discuss some typical examples for the photometric
properties of the {\sc COMARCS} atmospheres studied here. While in Fig.~\ref{cst_colt}
(J$-$K) and (H$-$K) are shown as a function of the effective
temperature, Fig.~\ref{cst_colcol} contains a two-colour diagram (J$-$H) versus
(H$-$K). All graphs cover the same sequences of models, which correspond to
different abundance combinations. In Fig.~\ref{cst_colcol} one can also find
results from the previous carbon star grid by \citet{2009A&A...503..913A} and
observations. These data will be discussed in the following sections. For
[Z/H]~=~0 and $-2$ we compare {\sc COMARCS} atmospheres with [O/Z]~=~+0.5 and
[O/Z]~=~+0.5 plus [N/Z]~=~+0.5 to similar ones having a constant value of
C/O or [C$-$O], where a scaled solar composition is assumed. In the case of
[Z/H]~=~$-1$ the graphs include a sequence with an oxygen enrichment of
[O/Z]~=~+0.25 at C/O~=~2.0, which was taken from the stellar evolution
calculations. It can be checked against models characterized by the primordial
[O/Z] of $-0.04$ and the same C/O, [C$-$O] or total amount of carbon.

In Fig.~\ref{cst_colt} one can see that the largest shifts in (J$-$K) exceeding
0.1~mag appear below 3100 to 3200~K and are due to the different values for the
overall metallicity. The stars get significantly bluer, if [Z/H] decreases. This
effect is a lot stronger than the variations caused by a temperature offset of
100~K or by the deviations of individual abundances investigated here. The
situation changes for the warmer models, where we find no clear trend with
metallicity and the corresponding shifts become much smaller. In the range below
3000 to 3100~K an enrichment of nitrogen by [N/Z]~=~+0.5 gives a moderate
reddening between about 0.03 and 0.06~mag, while it generates almost no or no
differences, if the stars are hotter. The {\sc COMARCS} atmospheres having a constant
C/O ratio get in general bluer, when the amount of free carbon atoms [C$-$O]
increases due to a higher [O/Z]. However, the corresponding shifts remain often
very small not exceeding 0.01 or 0.02~mag for the shown examples. Only in the
cooler models at solar metallicity and around 3000 to 3200~K at a lower [Z/H]
we find moderate deviations, which grow in the case of [O/Z]~=~+0.5 up to
about 0.05 or 0.06~mag. It should also be noted that the (J$-$K) colours for
the same [C$-$O] and different C/O are always quite similar with offsets of
less than 0.01 to 0.03~mag.

In the lower panel of Fig.~\ref{cst_colt} we observe also a strong influence
of the metallicity on the (H$-$K) colours of the {\sc COMARCS} models. However, there
is no clear trend and the corresponding differences are not larger than those
caused by the studied variations of C/O or [O/Z]. The shifts produced by an
enrichment of nitrogen remain in general quite small never exceeding 0.02~mag
for the shown cases with [N/Z]~=~+0.5. It is obvious that the amount of free
carbon atoms plays an important role for the predicted (H$-$K) colours. At all
of the covered effective temperatures and metallicities they become redder, if
[C$-$O] increases. Adopting [O/Z]~=~+0.5 and a constant C/O ratio the related
offsets can get larger than 0.05~mag corresponding to a model with identical
abundances, which is much more than 100~K cooler. At least for the higher
[Z/H] values the amount of oxygen has some additional influence due to the CO
lines. The colours turn bluer, when the carbon excess remains the same and
C/O increases. This effect is always weaker than the reddening caused by a
larger [C$-$O], and it almost disappears for the lowest metallicities.

The appearance of the {\sc COMARCS} atmospheres in the (J$-$H) versus (H$-$K) diagram
presented in Fig.~\ref{cst_colcol} reflects basically the temperature dependence
shown in the two panels of Fig.~\ref{cst_colt}, since (J$-$H) and (J$-$K) behave
in a similar way. For both of the last-mentioned colours we find a clear trend
that below 2800 to 3100~K cooler models become again bluer, which is much more
pronounced in metal-poor stars. This reversion can usually not be observed, because
the corresponding objects are reddened by dust. The issue was already discussed
in \citet{2009A&A...503..913A}. All or almost all carbon stars with effective
temperatures below 2800 to 3000~K have circumstellar shells generating significant
shifts of the spectral energy distributions, which is also confirmed by the
photometric measurements included in Fig.~\ref{cst_colcol}.

In Fig.~\ref{cst_colcol} one can see that the deviations of the involved colours
due to a higher nitrogen abundance of [N/Z]~=~+0.5 remain in general small
compared to other effects. At the same time, we get considerable shifts, if the
value of [C$-$O] changes. For the shown examples these variations are always
larger than the ones caused by a different C/O ratio at a constant carbon
excess. Nevertheless, also for the latter case we find significant offsets, when
the metallicity becomes higher. The role of [C$-$O] and [Z/H] as key parameters
for the photometric properties will be discussed in the next section. Although
those two quantities are very important for the spectra, their determination
based on colours alone is usually not possible. The different dominating
trends due to the abundances interact with each other, as one can see in
Fig.~\ref{cst_colcol}, and with uncertainties concerning the effective
temperatures, surface gravities, circumstellar reddening and deviations
from a hydrostatic structure.

\subsubsection{The Carbon Excess}
\label{sec_carex}

\begin{figure}
\includegraphics[width=9.9cm,clip,angle=0]{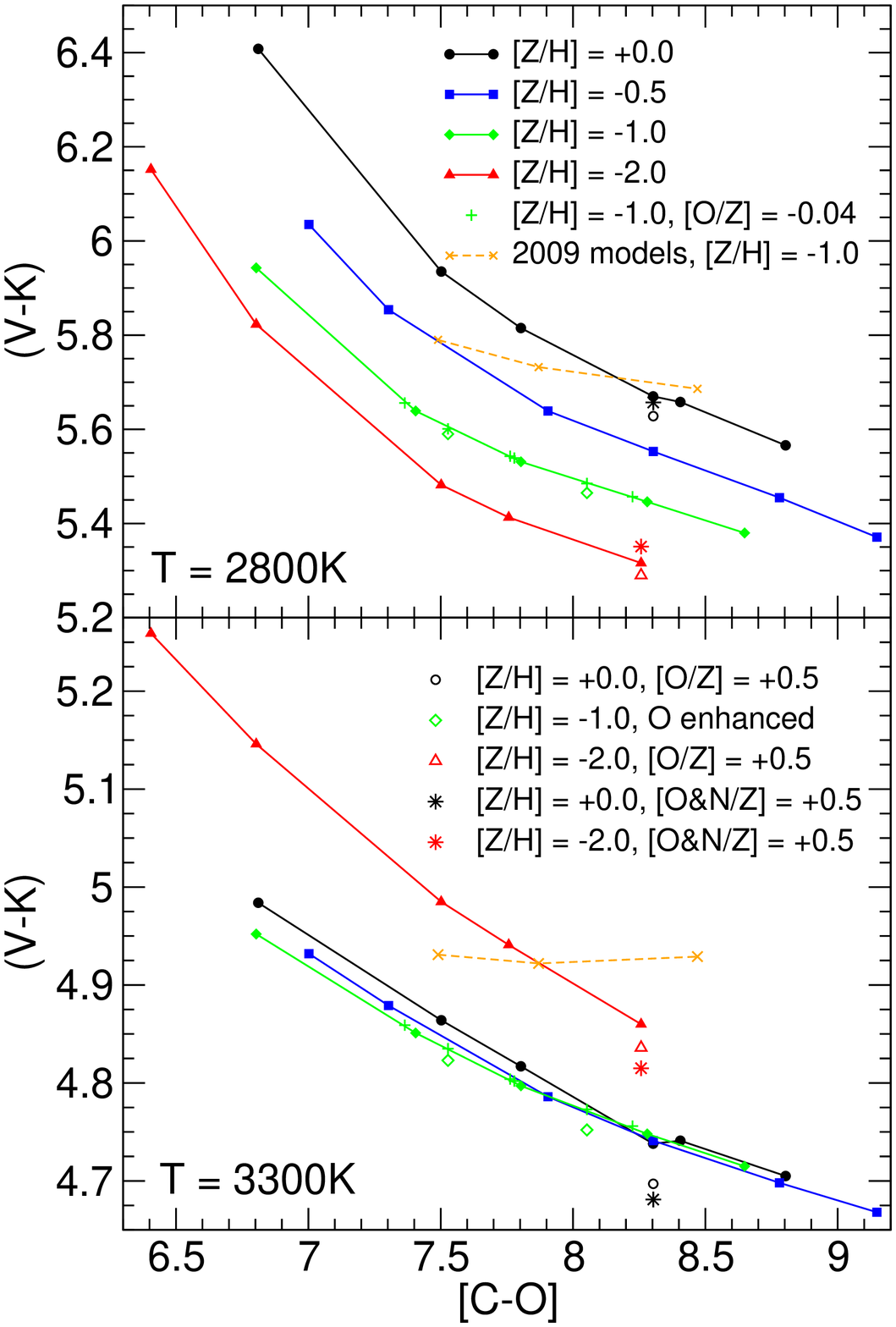}
\caption{The (V$-$K) colour as a function of [C$-$O] for {\sc COMARCS} models with
$\rm log(g~[cm/s^2]) = 0.0$ and one solar mass having $\rm T_{eff} = 2800~K$
(upper panel) and 3300~K (lower panel). All abundance combinations used in this
work and listed in Table~\ref{ari_comarcs} are included. The colours in the
plot correspond to the different metallicities [Z/H]: 0.0 black, $-0.5$ blue,
$-1.0$ green and $-2.0$ red. Filled symbols connected by lines represent the
sequences of models with standard composition where [O/Z] and [N/Z] remain
0.0. Green crosses mark results for [O/Z]~=~$-0.04$ at [Z/H]~=~$-1.0$. Open
symbols stand for an enrichment of oxygen: [O/Z]~=~0.5 at [Z/H]~=~0.0 and
$-2.0$, [O/Z]~=~0.12 or 0.25 at [Z/H]~=~$-1.0$. An asterisk corresponds to
an increased abundance of oxygen and nitrogen: [O/Z]~=~0.5, [N/Z]~=~0.5. We
have also included models with [Z/H]~=~$-1.0$ and two solar masses (orange)
taken from the previous {\sc COMARCS} carbon star grid by \citet{2009A&A...503..913A}.}
\label{cst_vk}
\end{figure}

\begin{figure}
\includegraphics[width=9.9cm,clip,angle=0]{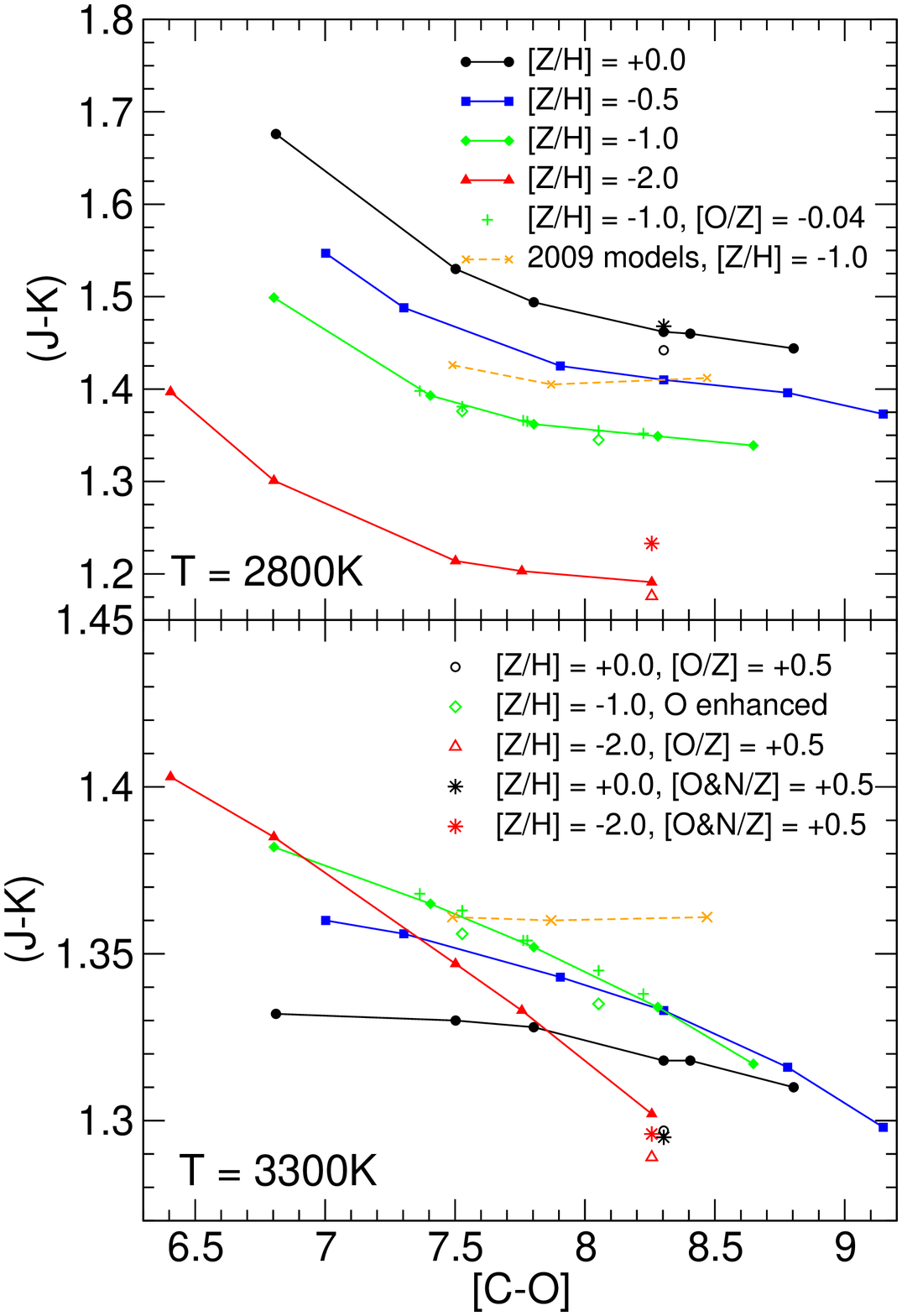}
\caption{The (J$-$K) colour as a function of [C$-$O]. The included {\sc COMARCS}
models and the corresponding symbols and line colours are the same as in
Fig.~\ref{cst_vk}.}
\label{cst_jk}
\end{figure}

\begin{figure}
\includegraphics[width=9.9cm,clip,angle=0]{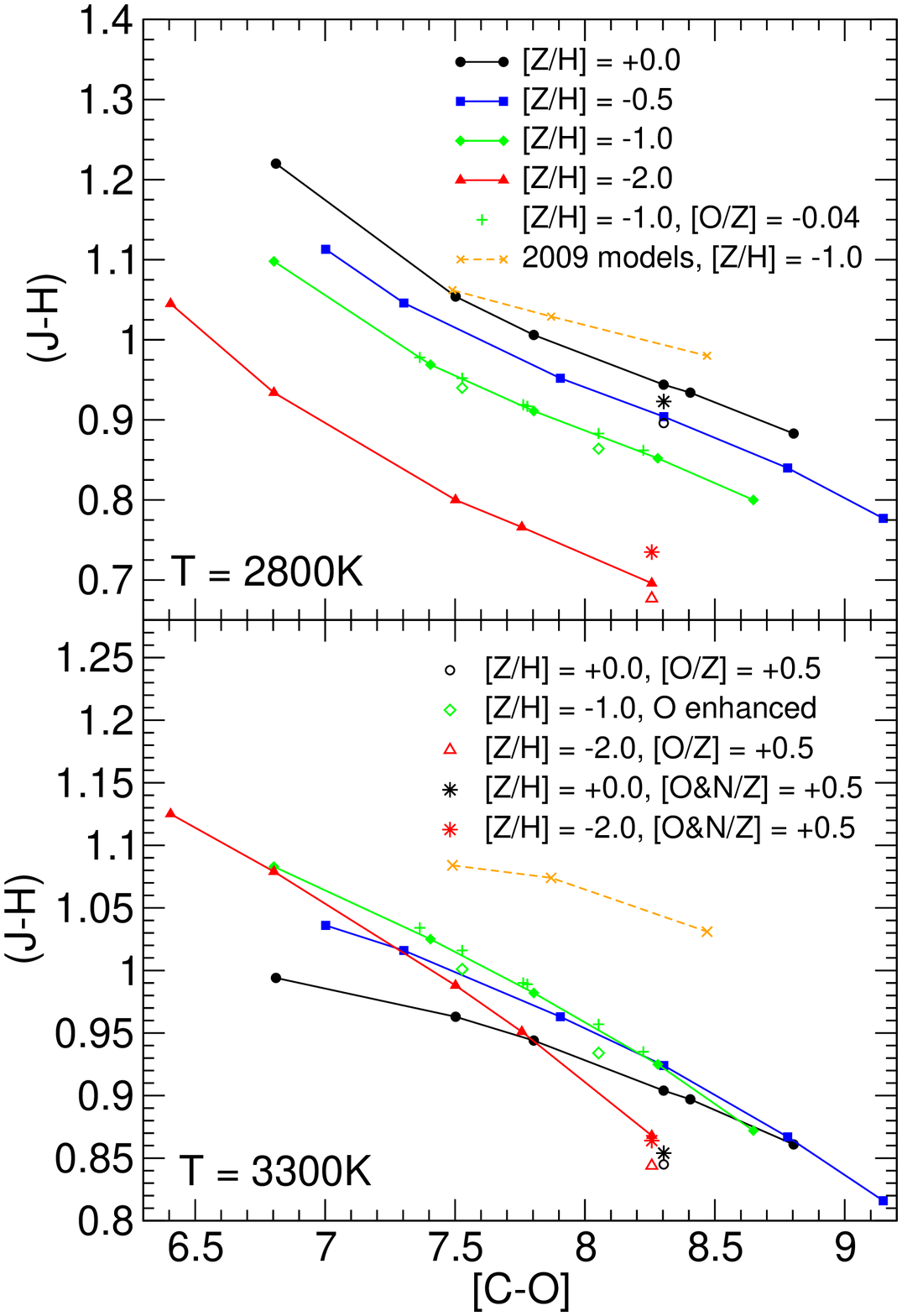}
\caption{The (J$-$H) colour as a function of [C$-$O]. The included {\sc COMARCS}
models and the corresponding symbols and line colours are the same as in
Fig.~\ref{cst_vk}.}
\label{cst_jh}
\end{figure}

\begin{figure}
\includegraphics[width=9.9cm,clip,angle=0]{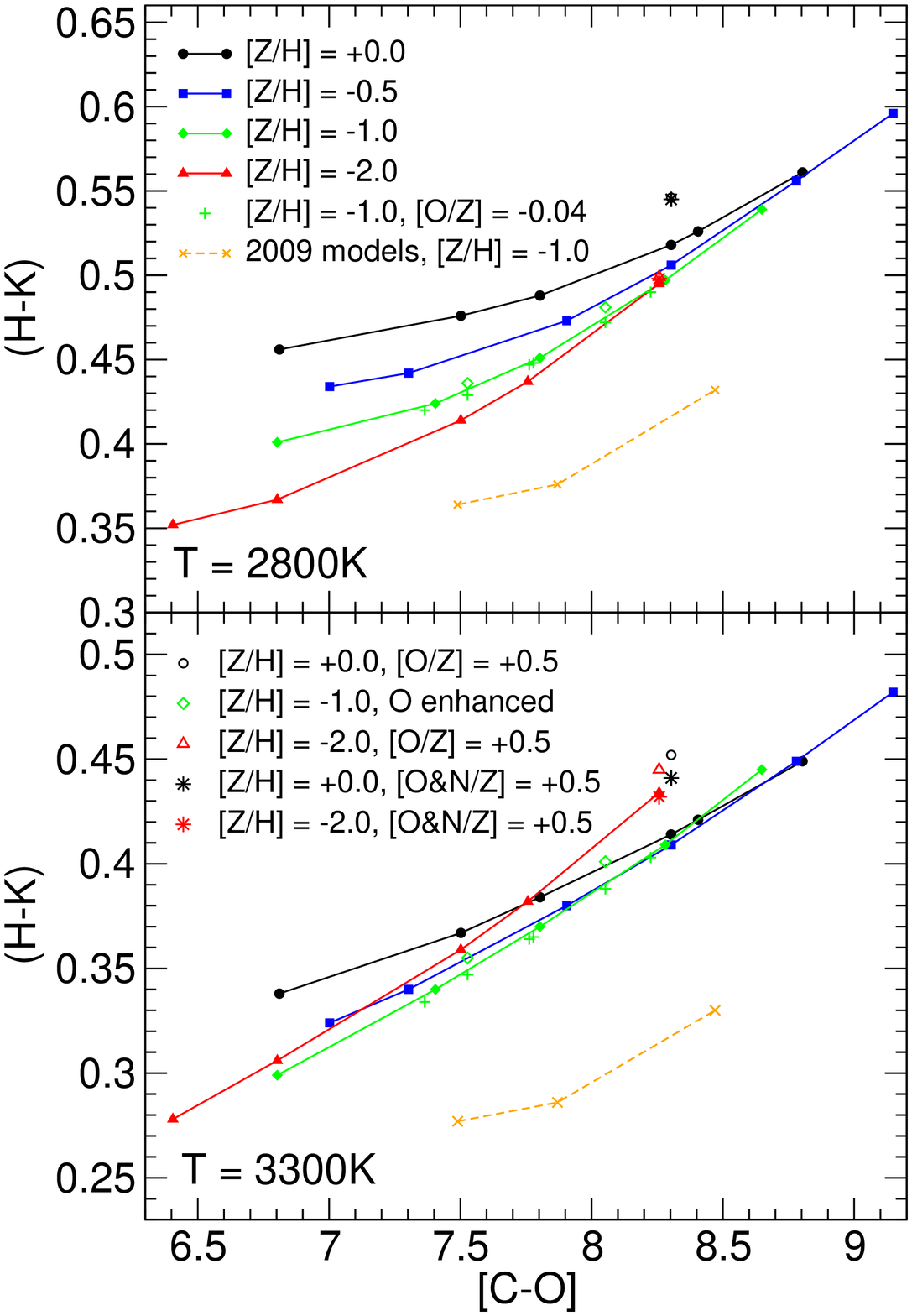}
\caption{The (H$-$K) colour as a function of [C$-$O]. The included {\sc COMARCS}
models and the corresponding symbols and line colours are the same as in
Fig.~\ref{cst_vk}.}
\label{cst_hk}
\end{figure}

One can conclude from the previous section that [C$-$O], which corresponds
in principle to the amount of free carbon atoms not bound in CO, is an important
parameter for the colours. In Figs.~\ref{cst_vk}, \ref{cst_jk}, \ref{cst_jh} and
\ref{cst_hk} we show (V$-$K), (J$-$K), (J$-$H) and (H$-$K) as a function of this
quantity. The plots cover all available abundance combinations listed in
Table~\ref{ari_comarcs} and two effective temperatures. While 2800~K (upper
panels) is typical for a cool carbon star, 3300~K (lower panels) represents
a warmer one having only weak bands due to polyatomic molecules. The different
colours in the graphs correspond to the included metallicities. Models with
scaled solar standard mixtures (see \ref{sec_sta}) are shown as filled symbols
connected by full lines. The positions of the various single characters
representing the other abundance combinations indicate the expected deviations
from these main relations, when the amount of oxygen or nitrogen is changed for
fixed values of [C$-$O] and [Z/H]. One can conclude for example that the
slightly smaller primordial [O/Z] of $-0.04$ taken from the stellar evolution
calculations (green crosses, see \ref{sec_evo}) causes almost no shifts. The
plots contain also models from the previous carbon star grid by
\citet{2009A&A...503..913A}, which will be discussed in the following sections
(orange, dashed lines).

For the models with [Z/H]~=~$-1.0$ and an oxygen enrichment of [O/Z]~=~0.12 or
0.25 taken from the stellar evolution calculations the deviations from the
sequences of results obtained using scaled solar abundances remain always
very small, if they are compared to the variations due to different values
of the carbon excess. At 2800~K also the {\sc COMARCS} atmospheres having [O/Z]~=~0.5
and [O/Z]~=~[N/Z]~=~0.5 are in most cases quite close to these standard
relations. However, a few exceptions with moderate changes exist. Shifts
of about 0.05 to 0.07~mag in (J$-$K) and (J$-$H) can be found for the
lowest metallicity, when the nitrogen abundance is increased. In addition, an
oxygen enrichment at [Z/H]~=~0.0 makes (H$-$K) almost 0.03~mag redder. For
3300~K the relative deviations of models with [O/Z]~=~0.5 and [O/Z]~=~[N/Z]~=~0.5
appear often larger, which is at least partly due to the smaller variation
of the colours as a function of the carbon excess. One should note that the
upper and lower panels in the figures have different scales. The absolute
offsets remain also for the warmer objects below about 0.03 to 0.06~mag. Compared
to the consequences of changing the effective temperature, surface gravity or
optical depth of a possible dust shell this is not much.

\begin{figure}
\includegraphics[width=9.9cm,clip,angle=0]{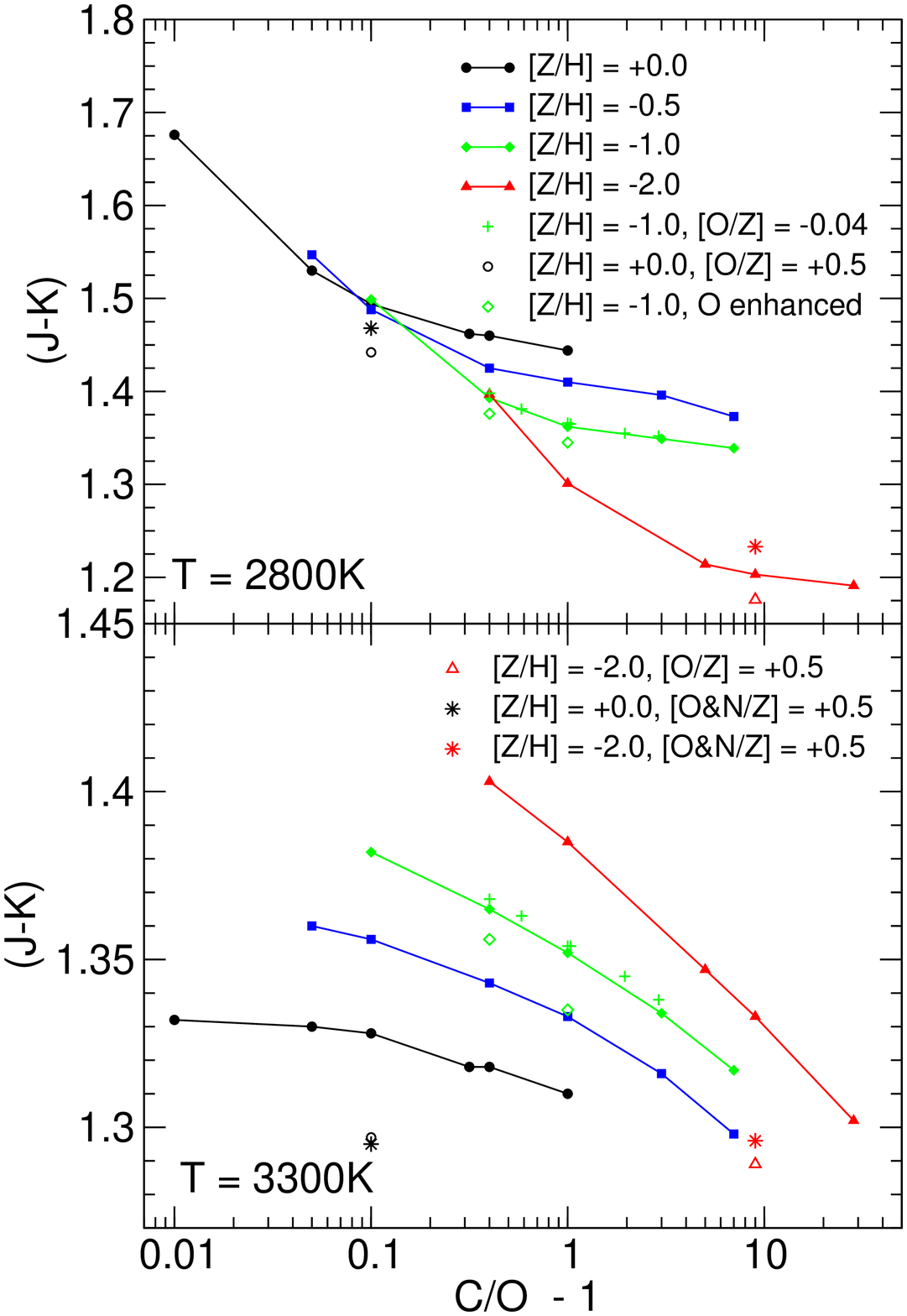}
\caption{The (J$-$K) colour as a function of C/O$-$1. The included {\sc COMARCS}
models and the corresponding symbols and line colours are the same as in
Fig.~\ref{cst_vk} except for the results from the previous carbon star
grid, which are not shown here.}
\label{cst_jkco}
\end{figure}

The effect of using C/O (in the plot C/O$-$1 for a better logarithmic
representation) instead of [C$-$O] as a key parameter can be seen in
Fig.~\ref{cst_jkco}, where we show (J$-$K) as an example. In the corresponding
two panels those {\sc COMARCS} atmospheres, which have already been covered by the
previous graphs featuring the carbon excess, are included keeping the same
symbols, lines and colours. When we compare the shifts caused by an oxygen
abundance deviating from the scaled solar value to the ones in
Fig.~\ref{cst_jk}, it is obvious that they are always larger in the
current diagram. Thus, using [C$-$O] and [Z/H] to characterize the chemical
mixture gives more accurate predictions for (J$-$K) than a similar combination
involving the C/O ratio. The same is normally also true for most of the other
colours.

\section{Discussion}
\label{sec_discussion}

In the following text we compare the photometric data obtained from the
{\sc COMARCS} models presented here to results based on the old (2009) carbon
star grid. In addition, the impact of the current investigation on the
predicted filter magnitudes for generating the observable properties of
synthetic stellar populations will be discussed. Finally, we study the
possibilities of measuring an enhancement of oxygen or nitrogen considering
the behaviour of spectra and colours described in the previous sections as
well as uncertainties concerning temperatures, surface gravities, dynamical
effects and dust shells.

\subsection{Comparison with the Previous Carbon Star Grid}

In Figs.~\ref{cst_vk} to \ref{cst_hk}, where colours are shown as a function
of [C$-$O], we have also included a sequence of models with [Z/H]~=~$-1.0$
taken from the previous carbon star grid by \citet{2009A&A...503..913A}. It
should be noted that these results were calculated assuming 2.0~M$_{\odot}$, while
all other {\sc COMARCS} atmospheres used in our plots have 1.0~M$_{\odot}$. For the
lower metallicities the old database does not contain the corresponding objects
with the smaller value. However, it has been demonstrated in
\citet{2009A&A...503..913A} that changes of the mass at constant effective
temperature and surface gravity cause only tiny colour shifts, which may in many
cases be neglected. Also the slightly different abundances adopted for the
solar mixture of the two grids will not generate significant deviations. Thus, we
can conclude that the considerable offsets between old and new {\sc COMARCS} models
are mainly due to the updates concerning the molecular opacities described in
Section~\ref{sec_mods}. These variations become often larger than the shifts
caused by a change of the metallicity or carbon excess.

As it was already shown by \citet{2009A&A...503..913A} the use of unscaled
C$_2$ data, which became the standard for the new {\sc COMARCS} models, results
in a much better agreement of synthetic and observed (H$-$K) or (J$-$H)
colours. This can be seen in Fig.~\ref{cst_colcol}, where we plot the two
quantities against each other. The included sequence from the old grid with
[Z/H]~=~0.0, 1.0~M$_{\odot}$ and C/O~=~1.10 remains for effective temperatures
around 2900 to 3200~K outside the region covered by the measured values. For
the corresponding new models having scaled solar abundances this is not the
case. The situation could change, if one assumes a considerable enrichment
of oxygen. However, we do not expect that all observed carbon stars have
[O/Z] values of about 0.3 or more, which may generate the necessary shifts. In
the range above 3000~K there are also no significant deviations due to a
different surface gravity, and objects with C/O~$\sim$~1.10 at solar metallicity
should be common. Thus, we can confirm that the new {\sc COMARCS} atmospheres allow a
better reproduction of the measured colours. A detailed discussion concerning the
observations is located in Section~\ref{sec_obs}.

In this context we also want to note that our preliminary tests with the new C$_2$
linelist from \citet{2018MNRAS.480.3397Y}, which is considered to be more complete
and accurate than the old one published by \citet{1974A&A....31..265Q}, confirm
our conclusion. The scaling used in the 2009 grid should not be applied, since it
makes the predicted spectra worse. The data of \citet{2018MNRAS.480.3397Y} will
become part of the standard opacity setup for the next generation of {\sc COMA}
calculations.

\subsection{Synthetic Colours for Stellar Populations}

In order to obtain absolute magnitudes or colours for isochrones and population
synthesis from the photometric tables of the {\sc COMARCS} grid, an interpolation in
metallicity, effective temperature, surface gravity and C/O ratio or [C$-$O] has
to be performed. The stellar mass is treated as a correction factor, which may
in the end be applied to the results. Our method was described in detail by
\citet{2009A&A...503..913A}. We use a linear interpolation taking [Z/H], log~(g)
and $\rm log(T_{eff})$ for the first three parameters. An example for such
calculations can be found in \citet{2017ApJ...835...77M}.

Since our systematic investigation of the effects caused by an increase of the
oxygen and nitrogen abundances exists only for a limited sample of carbon star
sequences with $\rm log(g~[cm/s^2]) = 0.0$ and one solar mass, an interpolation
including the corresponding parameters [O/Z] and [N/Z] is not possible. The
addition of two further dimensions would require a huge number of extra
models, which simply takes too much time. Thus, [O/Z] and [N/Z] have to be
considered as corrections applied to the results of the main computations. The
corresponding values can be derived using the data produced for the current
work. Such a treatment is similar to the one of the stellar mass.

However, as it has been shown in Section~\ref{sec_carex}, the deviations caused
by increased abundances of oxygen and nitrogen are in most cases small, if
[Z/H] plus [C$-$O] remain constant, and [O/Z] or [N/Z] does not exceed 0.5. The
uncertainties due to effective temperatures, surface gravities and reddening
by dust will usually be larger. Thus, for the moment, we assume that it is not
essential to introduce general corrections representing [O/Z] and [N/Z], when
isochrones or similar data are calculated. Only for a stronger enrichment of
oxygen and nitrogen with values above 0.5 the colour changes can become
quite significant. Nevertheless, we do not expect that this happens very often
in common carbon stars. In any case, it is important to use [C$-$O] instead of
C/O, which keeps especially in cooler objects the possible errors small. A comparison
of Figs.~\ref{cst_jk} and \ref{cst_jkco} shows the consequences of adopting those
two quantities as main parameters. It has already been said that with C/O the
deviations are always bigger, if one increases the abundance of oxygen. This is
expected, since the absolute number of carbon atoms not bound in CO dominates the
chemical equilibrium. A similar behaviour was found by \citet{2016MNRAS.457.3611A}
for very cool M giants, where [O$-$C] represents the important quantity.

\subsection{Models and Observations}
\label{sec_obs}

In this section we want to discuss, to what extent photometric and spectroscopic
observations can contribute to draw conclusions concerning carbon excess and
oxygen or nitrogen enrichment.

\subsubsection{Photometric Observations}

For the comparison of synthetic near infrared colours based on the new {\sc COMARCS}
model grid against various observational data in Fig.~\ref{cst_colcol} we
adopted photometry from a few sources in the literature, which shall be described
briefly in the following.

The first sample shown in the plot contains carbon stars in different parts of
both Magellanic Clouds as well as in our galaxy, for which J, H and K photometry
was obtained by \citet{1981ApJ...249..481C}. From their Tables~1 to 6 we could
adopt for all of the sources values already corrected for interstellar reddening. An
even more extended set of similar objects, namely C giants identified in the LMC, was
later presented by \citet{1996AJ....112.2607C}. The J, H and K data from their
Table~3 are also dereddened and included in our plot. According to the latter
authors, for both studies the same instrumental setup was applied corresponding
to the CIT/CTIO photometric system \citep{1982AJ.....87.1029E}. The measurements
were therefore converted to our standard colours following the equations given in
Table~I of \citet{1988PASP..100.1134B}. In addition, the plot contains the
collection of galactic C giants compiled from the literature by
\citet{2001A&A...369..178B} and used in \citet{2009A&A...503..913A}. According
to \citet{1997A&A...321..236K} those observations can be directly compared to
our synthetic photometry without any transformation. Finally, we show the colours
derived from the spectra obtained by \citet{2016A&A...589A..36G} for carbon stars
in the Magellanic Clouds and our galaxy. Because of the different locations of the
objects included in the plot we expect that they cover a large metallicity range.

The most obvious thing in Fig.~\ref{cst_colcol} is that the included hydrostatic
models are able to explain the colours of the bluer carbon stars, while they
fail completely to reproduce the observed objects with (H$-$K)~$>$~0.5. As
was already mentioned before, this behaviour is mostly caused by circumstellar
reddening. Calculations simulating the effect of a dust shell around a {\sc COMARCS}
atmosphere can usually predict the photometric properties of the cooler C giants
having massloss quite well \citep{2011A&A...532A..54S,2016MNRAS.462.1215N,
2018MNRAS.473.5492N}.

The various covered values of metallicity, carbon excess and oxygen enhancement
produce considerable differences in the two-colour diagram, which have already
been discussed in Section~\ref{sec_phot}. We concluded there that it is difficult
or impossible to obtain detailed information concerning the abundances from
photometric indices alone, since the corresponding effects interact with each
other and with uncertainties related to stellar parameters and circumstellar
reddening. The situation might improve a bit, if more and narrower filters are
used. However, even from Fig.~\ref{cst_colcol} one may deduce some basic
results. It is for example obvious that the warmest carbon stars with effective
temperatures above 3600~K can only be described by models having a low
metallicity. This conclusion will not change, if we consider colour shifts
due to a possible deviation of the surface gravities from our standard
value. It agrees also with the fact that the corresponding objects are
usually located in the Magellanic Clouds. In addition, for the warmer
stars having no significant massloss the plot favours compositions, where
the carbon excess remains small becoming mostly not much larger than 8.0. On
the other hand, in order to get winds driven by dust, higher [C$-$O] values
are needed in the cooler giants, since enough material has to be available
\citep{2014A&A...566A..95E,2010A&A...509A..14M}. Finally, we find no observed
objects in the region, where the models predict the location of the sources
with solar metallicity and a strong enhancement of oxygen. Such carbon stars
may still exist, if they are obscured by a circumstellar shell.

\subsubsection{Spectroscopic Observations}

Regarding the discussion of several selected low resolution spectra in
Section~\ref{sec_specl}, it turns out that it will be very difficult or
impossible to deduce quantities like [C$-$O], [O/Z] and [N/Z] from such
data. Clear overall trends with the carbon excess or an enhancement of
nitrogen appear, if the fluxes are normalized relative to the
continuum. However, due to the presence of many weak overlapping
absorption lines this can never be measured in observed cool stars. In
the unscaled spectra the effects of the abundance changes investigated
here remain much less pronounced, which is at least partly caused by
the strong influence of the molecular opacities on the atmospheric
temperature-pressure structures. In addition, the relation between the
intensity of certain features and the chemical composition may become
complicated. For example, more nitrogen does not always result in
deeper CN bands. This behaviour is also due to the saturation of
stronger lines combined with a depression of the surrounding continuum
by a large number of weak overlapping transitions, which are often
at least partly produced by the same species. In connection with
possible uncertainties concerning effective temperatures, surface
gravities, reddening by circumstellar dust and the flux calibration
of the observations, the described conditions make it very difficult
to obtain any information about abundances from low resolution spectra
of carbon stars.

\begin{figure}
\includegraphics[width=7.0cm,clip,angle=270]{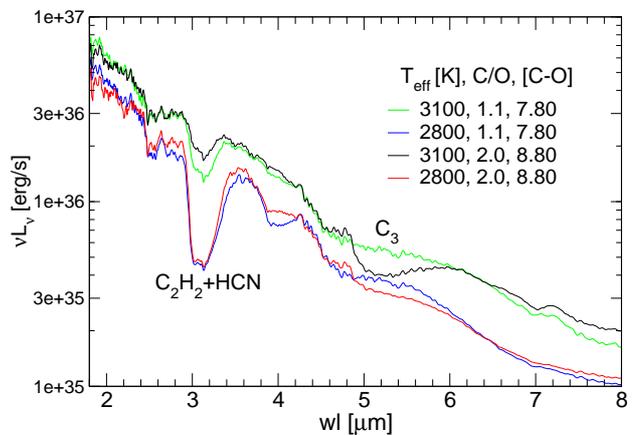}
\caption{Low resolution R~=~200 infrared spectra computed from {\sc COMARCS}
models with $\rm log(g~[cm/s^2]) = 0.0$, solar mass and metallicity. We show
the effect of increasing C/O from 1.1 to 2.0 for $\rm T_{eff} = 2800~K$ and
3100~K\@. The standard composition with [O/Z]~=~0 and [N/Z]~=~0 is assumed.
The strong features of C$_2$H$_2$ plus HCN at 3.2~$\mu$m and C$_3$ around
5~$\mu$m are marked.}
\label{cst_c3spec}
\end{figure}

A possible exception is the broad C$_3$ band around 5~$\mu$m, which could
be used to determine [C$-$O], since the corresponding molecular abundance
depends extremely on the amount of free carbon atoms
\citep{2000A&A...356..253J}. In Fig.~\ref{cst_c3spec} we show for two
effective temperatures (2800 \& 3100~K), how its intensity increases
with the C/O ratio. Also in this case a weakening of the feature due
to saturation combined with a depression of the surrounding flux level
appears, if the stars become cooler than about 3000~K\@. The {\sc COMARCS}
atmosphere with C/O~=~2.0 and 2800~K, which has by far the largest C$_3$
absorption in the plot, does not generate the most pronounced bump at
5~$\mu$m. In addition, the available hydrostatic and dynamic models do
not manage to reproduce the depth and shape of the strongest C$_3$ bands
correctly, which may be caused by a problem with the opacity data (independent
of the chemical data discussed in \citet{2000A&A...356..253J}).

For high resolution spectra the situation is much better, if one
chooses a range, where the quasi-continuous absorption of the weak
molecular lines does not get too big. The data shown in Section~\ref{sec_spech}
represent a good example. When saturated transitions and blends are
excluded, we see a clear correlation between the intensity of features
and the involved chemical abundances. This is an important condition for
measuring the carbon excess and an enrichment of oxygen or nitrogen using
lines of C$_2$, CN, HCN, CO and other species. However, in order to evaluate
the effect of a changed composition correctly, the treatment of molecular
opacities for the generation of models and synthetic spectra has to be
consistent.

\begin{figure}
\includegraphics[width=8.5cm,clip,angle=0]{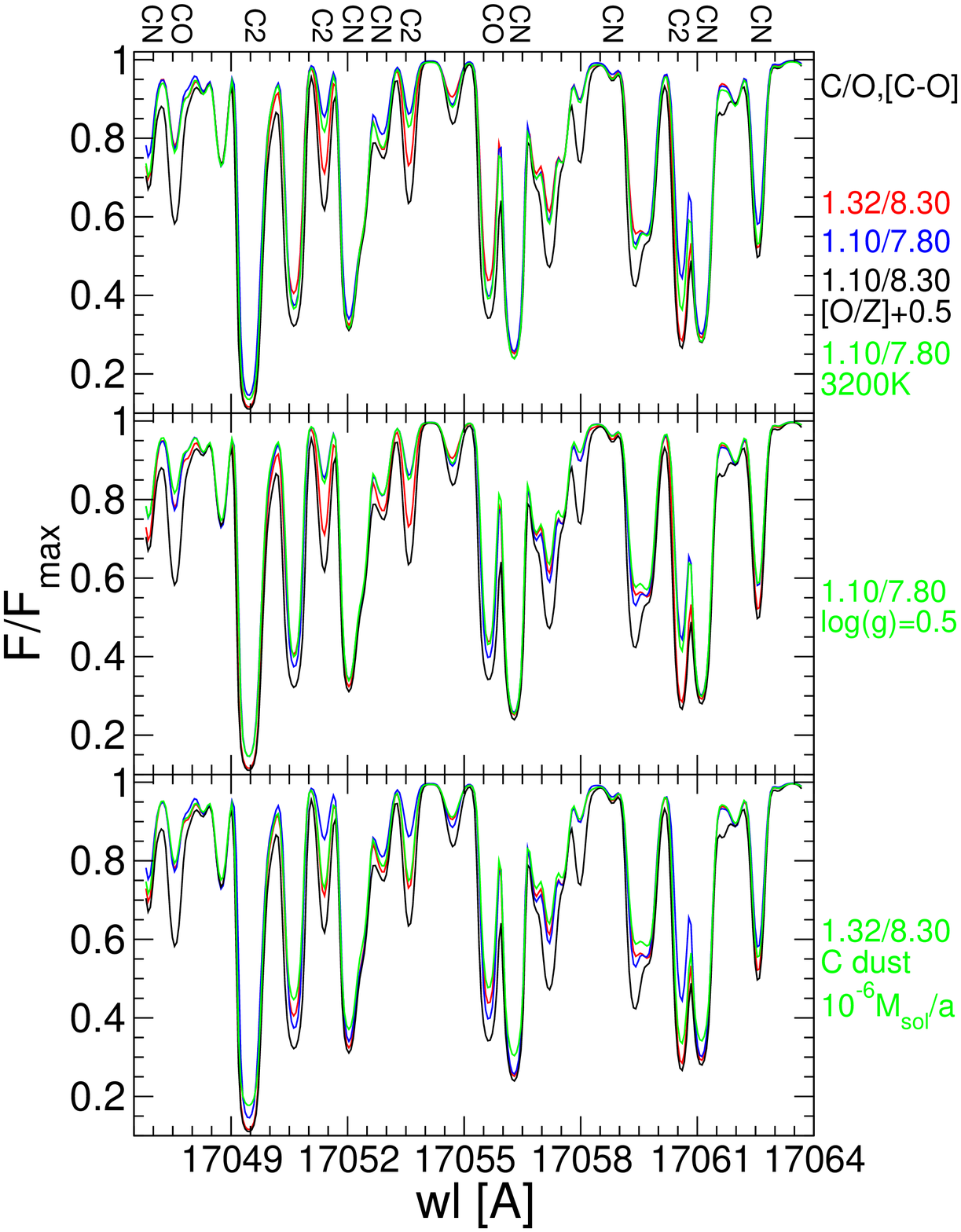}
\caption{High resolution R~=~300000 spectra based on {\sc COMARCS} models with
$\rm T_{eff} = 3300~K$, $\rm log(g~[cm/s^2]) = 0.0$, solar mass and metallicity
having [O/Z]~=~+0.5 (black) and the standard composition ([O/Z]~=~0) with
a constant C/O~=~1.10 (blue) or [C$-$O]~=~8.30 (red). They are compared to
calculations (green) for [O/Z]~=~0 and C/O~=~1.10, where $\rm T_{eff}$ was set
to 3200~K (upper panel) or $\rm log(g~[cm/s^2])$ to +0.5 (central panel). In
addition, we show results with [O/Z]~=~0 and [C$-$O]~=~8.30, which have been
obtained assuming an outer carbon dust shell computed for a massloss rate of
$10^{-6}$~M$_{\odot}$/a (green, lower panel). The data are normalized relative
to their point with the largest flux. Absorption features produced by a single
molecular species have been marked.}
\label{cst_hires3}
\end{figure}

In Fig.~\ref{cst_hires3} we study the consequences of a possible dust shell
and of uncertainties concerning the effective temperature or surface
gravity. They are compared to some of the changes caused by abundance
differences, which were discussed in Section~\ref{sec_spech}. The plot
includes three models shown already in Fig.~\ref{cst_hires} covering
the same spectral range. For a star having $\rm T_{eff} = 3300~K$,
$\rm log(g~[cm/s^2]) = 0.0$, solar mass and metallicity we present
a result with [O/Z]~=~+0.5 and two calculations, where the standard
composition ([O/Z]~=~0) is assumed keeping C/O~=~1.10 or [C$-$O]~=~8.30
constant. The variation due to the chemical mixture can then be compared to
the consequences of decreasing $\rm T_{eff}$ by 100~K and increasing
$\rm log(g~[cm/s^2])$ by 0.5. This are typical values of the uncertainties
expected for the parameters of cool carbon stars. In order to be able to
estimate the influence on the determination of abundances, we have marked
all features, which are produced by a single molecular species (no blend).

In the lower panel of Fig.~\ref{cst_hires3} we show the effect of a dust
shell consisting of amorphous carbon, which was computed using the spherical
stationary wind code by \citet{2013MNRAS.434.2390N,2016MNRAS.462.1215N}
assuming a massloss rate of $10^{-6}$~M$_{\odot}$/a. The corresponding opacity
data were taken from \citet{1991ApJ...377..526R}. A {\sc COMARCS} model with the
same stellar parameters as all the others in the plot, [O/Z]~=~0 and
[C$-$O]~=~8.30 has been adopted as the central radiation source. The higher
carbon excess resulting in C/O~=~1.32 was chosen for the comparison, since
it favours the formation of dust. We assumed an initial outflow velocity of
4~km/s, a primordial grain size of $10^{-7}$~cm and a normalized (relative to H)
seed particle density of $1.058 \cdot 10^{-12}$. The contribution of the gas in the
shell was neglected in the radiative transfer calculations carried out with
the {\sc COMA} code, which were also used to determine the temperatures of the
solid material in the wind. We have adopted the chemical composition of the
{\sc COMARCS} model for all computations.

One can see that the studied changes of effective temperature and surface
gravity have an impact on some of the absorption features. However, especially
for the CO and C$_2$ lines, which are not saturated, the shown variations due
to the abundances get much stronger. Thus, considering only the discussed
uncertainties of the stellar parameters, it should be possible to measure
differences of [C$-$O] or [O/Z], if they exceed values of about 0.2 to 0.3.

In our example also the attenuation of the absorption features caused by
circumstellar dust, which affects mainly the stronger lines, is much weaker
than some of the variations due to the abundances. Nevertheless, for higher
massloss rates or [C$-$O] values and in other wavelength ranges the influence of
such a shell could increase significantly. In extreme cases the spectral signature
of the central atmosphere may disappear completely. The amount of dust present
around an observed carbon star can usually be estimated from the overall
energy distribution deduced from the combination of different photometric
measurements \citep{2011A&A...532A..54S,2012A&A...543A..36G,2012A&A...537A.105G,
2012ApJ...753...71R,2018MNRAS.473.5492N}.

We conclude from our experiment that in stars with weak or moderate reddening
due to circumstellar dust ((J$-$K) up to about 3) the carbon excess and a
possible enrichment of oxygen can be measured. The same applies to the nitrogen
abundance, which may be determined using the CN lines. However, it was always
assumed that the inner atmospheres behave like hydrostatic structures with
radial symmetry. As we have already stated before, this approach will not
work, if the environment is dominated by pulsation, shock waves and
massloss. Such conditions appear mainly for the very cool carbon stars. In
their dynamic atmospheres deviations from hydrostatic equilibrium and the
related velocity fields change the intensities and profiles of lines, which
may even result in emission components. Examples for the effect on high
resolution spectra of M and C giants can be found in \citet{1992A&A...253..203S},
\citet{1996A&A...307..481B}, \citet{2000A&A...359..651W}, \citet{2010A&A...514A..35N},
\citet{2010A&A...517A...6L} or \citet{2017A&A...606A...6L}. It is obvious that
for those objects an abundance determination will be difficult or impossible
\citep{2007MNRAS.378.1089M,2014A&A...567A.143L}.

Thus, the carbon excess and an enrichment of oxygen or nitrogen can only be
measured in the atmospheres of giants, when pulsation and massloss do not
become too strong. On the AGB this corresponds usually to the less evolved
and warmer stars. Additional uncertainties affecting also those objects may
be due to deviations from spherical symmetry created by convection. For
M supergiants the phenomenon and its consequences concerning high resolution
spectra were investigated by \citet{2011A&A...535A..22C}. In the case of
extended carbon stars one might expect a similar behaviour. Finally, errors
affecting line intensities and positions can cause further problems
\citep{2005hris.conf..303A}. First of all, this will have an impact on
the identification of features produced by a single transition, since
many of them are blends involving different species. A large part of the
spectral signatures appears only in cool carbon stars, which prevents
a calibration using warmer and less complicated atmospheres. However, the
situation improves, as better molecular line data become available
\citep{2012MNRAS.425...21T,2014MNRAS.437.1828B,2014ApJS..210...23B,
2014ApJS..214...26S,2014A&A...571A..47M,2017JQSRT.201...94L,
2018JQSRT.204...42C,2018MNRAS.480.3397Y}. Unfortunately, not all of the
new lists are complete in the temperature range important for the
photospheres of red giants, which is a condition for their use in the
calculation of model structures and broad stellar spectra.

\section{Conclusions}
\label{sec_conclusions}

In addition to our standard C star grid with scaled solar chemical mixtures, where
only the amount of carbon has been increased, we have calculated sequences of
hydrostatic {\sc COMARCS} atmospheres assuming higher abundances of oxygen and nitrogen
for a constant C/O, [C$-$O] or $\rm \varepsilon_C$. This investigation remains
restricted to models with $\rm log(g~[cm/s^2]) = 0.0$ and 1.0~M$_{\odot}$. Overall
metallicities between [Z/H]~=~0.0 and $-2.0$ are covered. The [C$-$O] values range
from 6.41 to 9.15. The considered abundance changes may be caused by a combination
of internal nuclear processes and convective mixing events appearing in evolved
cool giants. They are expected following the predictions of stellar models and
observations of post-AGB stars. Our study comprises [O/Z] and [N/Z] values up to
+0.5.

Based on the atmospheric structures we have computed synthetic R~=~10000 OS and
convolved R~=~200 low resolution spectra as well as photometric data, which are
all available at \url{http://stev.oapd.inaf.it/atm}.\footnote{Or also at
\url{http://starkey.astro.unipd.it/atm}.} The observable properties
were derived using exactly the same setup concerning opacities and abundances as
in the models. Such a consistent treatment is important in order to obtain reliable
results. For some of the {\sc COMARCS} atmospheres we have also calculated high resolution
spectra allowing a study of individual lines.\footnote{Due to the discussed improvements
we recommend that the users of our synthetic spectra and photometry for carbon stars
should always take the data presented in this work and \citet{2016MNRAS.457.3611A}
instead of the old ones from \citet{2009A&A...503..913A}, if new results are available
for the selected parameters.}

We find that it is difficult or not possible to get information about the abundances
of oxygen and nitrogen as well as the carbon excess from our R~=~200 spectra. When the
lines are not resolved, changes of the visible continuum due to the influence of
molecular opacities on the atmospheric structures and due to many weak overlapping
transitions, plus saturation in regions with strong absorption affect the appearance
of the features. Their intensity may remain constant or show a complex behaviour, if
[O/Z], [N/Z] and [C$-$O] increase. This is combined with uncertainties concerning
the flux calibration of the observations, the stellar parameters or the impact
of dust absorption, pulsation and massloss. It has already been mentioned that
the OS spectra, which have a higher resolution, can not be compared directly to
measurements.

For the high resolution spectra we have demonstrated that it is in principle
possible to measure an enrichment of oxygen plus nitrogen as well as the carbon
excess, if one chooses a region, where at least some of the more intense
CN, C$_2$ and CO lines are not blended or saturated and the influence of weak
overlapping molecular transitions on the observed continuum remains limited. The
changes produced by the typical uncertainties of the stellar parameters or a dust
shell corresponding to a moderate massloss are in general smaller than the
differences due to the abundance variations investigated here. However, as has
been mentioned, when a star shows stronger pulsations and winds, any analysis
based on hydrostatic models will not work. In this case the optical depth of the
envelope around the central object may become very large and dynamic processes
like shocks or episodic outflow dominate the structures affecting line intensities
and profiles. Such problems arise mainly for the coolest giants with
$\rm T_{eff} < 3000~K$. Another possible source of errors are the molecular
opacity data, which have an impact on the predicted strength of the measured
features, the appearance of blends and the influence of weak transitions on the
continuum. The corresponding uncertainties also grow for lower temperatures. At
high resolution it is again necessary to do the radiative transfer for the
calculation of models and synthetic spectra in a consistent way.

For the photometric data circumstellar dust causes the strongest changes. Many
of the observed carbon stars are in general much redder than all of the
hydrostatic models. This affects again especially the coolest giants with
$\rm T_{eff} < 3000~K$. Such objects can be described by adding a synthetic
external shell to the central atmosphere or by a complete dynamic calculation
including pulsation and massloss.

At a constant overall metallicity the carbon excess is the most important
abundance parameter for the colours. We have shown that using [C$-$O] instead
of C/O gives in general much better results, if one neglects a possible
deviation of the amount of oxygen from a scaled solar mixture. As has
been mentioned, the {\sc COMARCS} database does not (and will not) cover variations
of [O/Z] and [N/Z] with complete grids. However, like the stellar mass they
may be included by correction terms derived from the existing model sequences
and applied to the colours predicted for a certain combination of effective
temperature, surface gravity, metallicity and carbon excess. It is not
possible to determine the corresponding abundances from photometric
measurements alone, since problems similar to those discussed for the
low resolution spectra will appear. Nevertheless, even from the simple
(J$-$H) versus (H$-$K) diagram shown in this work one can get some
information. For example, it is obvious that the warmest carbon giants
hotter than 3600~K are metal-poor. We may also exclude the frequent appearance
of a strong enhancement of oxygen with [O/Z] around 0.5 in stars having a
higher [Z/H] close to zero and a weak massloss.

In the future it could be useful to compute grids of dynamic models including
pulsation, dust formation and massloss with an enrichment of oxygen or
nitrogen in order to cover also the cooler carbon giants. But even then
it will remain very difficult to estimate the elemental abundances in these
stars. We expect improvements concerning the hydrostatic atmospheres and
synthetic spectra mainly from a lot of new molecular opacity data, which
are already or may soon be available. The next versions of {\sc COMA} and {\sc COMARCS}
will include them. Examples relevant for carbon stars are CN, CH, C$_2$, C$_2$H$_2$
or C$_3$.

\section*{Acknowledgements}

This work was mainly supported by the ERC Consolidator Grant funding scheme
({\em project STARKEY}, G.A. n.~615604). We thank Marco Dussin for helping us with
the electronic publication of the data. AN acknowledges the support of the Centre National
d'{\'E}tudes Spatiales (CNES). We thank Katy Chubb for her support concerning the text.

\bibliographystyle{mnras}
\bibliography{csterne}

\bsp
\label{lastpage}
\end{document}